# Influence of Pore Surface Chemistry on the Rotational Dynamics of Nanoconfined Water

Benjamin Malfait[1], Aicha Jani[1], Jakob Benedikt Mietner[2], Ronan Lefort[1], Patrick Huber[3,4,5,*], Michael Fröba[3,*], and Denis Morineau[1,*]

[1]Institute of Physics of Rennes, CNRS-University of Rennes 1, UMR 6251, F-35042 Rennes, France

[2]Institute of Inorganic and Applied Chemistry, University of Hamburg, 20146 Hamburg, Germany

[3]Institute for Materials and X-ray Physics, Hamburg University of Technology, 21073 Hamburg, Germany

[4]Centre for X-Ray and Nano Science CXNS, Deutsches Elektronen-Synchrotron DESY, 22603 Hamburg, Germany

[5]Centre for Hybrid Nanostructures CHyN, Hamburg University, 22607 Hamburg, Germany

*Email:patrick.huber@tuhh.de

*Email:froeba@chemie.uni-hamburg.de

*Email:denis.morineau@univ-rennes1.fr

ABSTRACT: We have investigated the dynamics of water confined in mesostructured porous silicas (SBA-15, MCM-41) and four periodic mesoporous organosilicas (PMOs) by dielectric relaxation spectroscopy. The influence of water-surface interaction has been controlled by the carefully designed surface chemistry of PMOs that involved organic bridges connecting silica moieties with different repetition lengths, hydrophilicity and H-bonding capability. Relaxation processes attributed to the rotational motions of non-freezable water located in the vicinity of the pore surface were studied in the temperature range from 140 K to 225 K. Two distinct situations were achieved depending on the hydration level: at low relative humidity (33% RH), water formed a non-freezable layer adsorbed on the pore surface. At 75% RH, water formed an interfacial liquid layer sandwiched between the pore surface and the ice crystallized in the pore center. In the two cases, the study revealed different water dynamics and different dependence on the surface chemistry. We infer that these findings illustrate the respective importance of water-water and water-surface interactions in determining the dynamics of the interfacial liquid-like water and the adsorbed water molecules, as well as the nature of the different H-bonding sites present on the pore surface.

# 1. INTRODUCTION

Water is undoubtedly one of the most important substances due to its central role as a solvent in an extensive number of natural and industrial processes. In most frequent situations, water is found as spatially confined or in an interfacial state rather than forming a bulk phase. This is the case for a number of geological materials (zeolites, clays…) and related applications for environmental remediation and waste water treatment.[1] Confined water is also crucial in energy technology, including proton-exchange membrane fuel cells,[2,3] and many other perspectives for water–energy technologies that were opened by the emergence of nanofluidics.[4] In biological systems too, water molecules interacting closely with the surface of biomolecules form an hydration layer (0.4 – 0.8 nm),[5] that differs from bulk water,[6] and can be considered as a confined phase.[7] Water dynamics in this layer is partly determined by the interactions between the water and the functional groups of the macromolecule.[6,8–10]

From a fundamental point of view, confining water at the nanoscale in prototypical porous solids has turned out to be particularly adequate in order to better understand the unusual behavior of interfacial water. For pore diameter $d < 2.5$ nm,[11] crystallization of confined water is suppressed by inhibition of the nucleation process,[12,13] which was considered as an opportunity to study anomalous physical properties of liquid water in a wide temperature range.[7,14] Among several types of confinement, including clay,[15] graphite oxide,[16] zeolite,[17] or porous silica glasses,[18,19] the mesostructured SBA-15[20] and MCM-41[21] are particularly suited hosts due to their well-defined porous geometry formed by ordered cylindrical channels.[22,23] Dynamics of confined supercooled water in MCM-41 has been studied with various experimental techniques as dielectric relaxation spectroscopy (DRS),[21] nuclear magnetic resonance (NMR),[24] and quasi-elastic neutron scattering (QENS).[25]

The impact of the pore size, surface chemistry or filling fraction has been reviewed in the literature and remains a matter of discussion.[7,20] In the high temperature region (ca. above 240 K), the main structural relaxation $\alpha$-process, varies markedly with the size of confinement. Notably, when decreasing the pore diameter, the temperature dependence of the relaxation time evolves from a non-Arrhenius towards a more Arrhenius-like behavior.[26] At lower temperature, a crossover was detected at around 225 K from non-Arrhenius behavior at high temperature to Arrhenius-like behavior at low temperature. This feature was detected for different confinement sizes, also corresponding to circumstances where water was present in different physical states, namely (i) liquid water coexisting with ice for SBA-15 (pore size 6 nm),[20] (ii) liquid with partial crystallization occurring in the crossover region for MCM-41 (pore size from 2.4 to 3.6 nm),[20,27–29] and (iii) entirely non-freezing liquid for pore size smaller than 2.4 nm.[29–31] In the low temperature regime, the reorientation dynamics of water molecules hardly depends on the pore size, suggesting that it reflects an intrinsic property of interfacial water.[7,20,28,32] This can be rationalized when noting that for sufficiently large pores, ice formation further restricts the interfacial volume accessible to the liquid phase.[26,27] In this situation, the geometrical confinement is determined by the thickness of the interfacial layer itself and it is no longer related to the pore size. Concerning the surface chemistry, $^2$H NMR experiments revealed similar reorientation dynamics of water molecules confined in modified and unmodified MCM-41 (with equivalent pore size),[26] which supports that it reflects a universal character of the relaxation dynamics of water. However, this observation was shown to depend strongly on the degree of pore filling. For partially filled pores, the reorientation dynamics of the adsorbed water layer is significantly faster (three decades) than the interfacial water when the pore center is entirely filled.[27,29] This demonstrates the critical importance of the nature of interfacial interactions (water-vapor, water-water and water pores), which are promoted when considering partially or capillary filled samples.[7]

The emerging question from all these results concerns the influence of the surface chemistry of the pore on the dynamics of surface water. In order to extent current knowledge, which has so far been based on a few studies on grafted silicas, we are contemplating new opportunities offered by the molecular scale imprint of the water-surface interaction. Periodic Mesoporous Organosilicas (PMOs), are particularly well-suited, though barely used in water studies so far. Their pore wall is formed by a periodic repetition of inorganic and organic bridging units, with tunable hydrophilicity and H-bonding tendency.[33] Unlike post-synthesis surface grafted porous silicas,[34] PMOs allow a stoichiometric control of the periodically alternating surface chemistry along the pore channel (i.e. one organic bridge per silica inorganic unit).[35]

Multidimensional solid-state NMR study has shown that the specific surface chemistry of PMOs caused different spatial arrangements of the interfacial water.[35] Strong correlations between water and the surface of the pore were observed in the vicinity of inorganic units that present H-bonding silanols groups. This was also the case near hydrophilic organic bridges when they comprise amino groups that act as alternative H-bonding sites. On the contrary, a depression of the interfacial layer of water around hydrophobic organic bridging units was indicated by the absence of spatial correlation.[35] Very recently, QENS experiments were carried out to investigate the dynamics of liquid water confined in PMOs, above the melting point, in the relatively high temperature range of 243 to 300 K.[36] This study concluded on the spatial heterogeneity of the translational dynamics across the pore diameter. Moderate confinement effects on the translational dynamics were observed for the fraction of most mobile molecules, which were likely located in the pore center, and thus were not affected by the nature of the pore surface. However, the interaction strength determined the longtime tail of the time dependence of the translational dynamics, corresponding to interfacial water molecules close to the surface wall.[36]

The above-mentioned observations made for water-filled PMOs underline the interest in focusing on the dynamics of only the few molecules layers that interact with the pore surface, and also in extending the experimental dynamical range to longer time. In the present study, we have selected two different experimental conditions that fulfill this requirement: (i) by loading the pores at 33% RH below the capillary condensation pressure to ensure that water is adsorbed on the surface, leaving the pore center empty, and (ii) by decreasing the temperature of saturated pores at 75% RH to allow crystallization of the pore center. Indeed, in the latter case, it is expected that an interfacial layer of liquid water remains, with a thickness of about 0.6 nm,[37] which is also comparable with the hydration layer (0.4 – 0.8 nm).[6] We have applied dielectric relaxation spectroscopy, taking advantage of its higher sensibility and wider accessible timescale (microsecond to second) to study the low temperature and longtime rotational dynamics of interfacial or monolayer water molecules.

## 2. RESULTS AND DISCUSSIONS

**2.1. Samples and Methods.** We used four different periodic mesoporous organosilicas (PMOs) with different organic bridging groups: divinylbenzene (DVB), divinylaniline (DVA), benzene (B) and biphenyl (BP), with mean pore diameter $D$ in the range 3.5 to 4.1 nm. Comparing results obtained with these PMOs allows studying both the effect of hydrophobicity of the organic bridge (DVA-PMO vs DVB-PMO) and repetition distances of silica/organic units along the pore axis that is related to the length of the organic unit (B-PMO vs BP-PMO), as illustrated in Figure 1. These materials were complemented by purely siliceous matrices with comparable mesoporous geometry (MCM-41, $D$ = 3.65 nm) or larger pore size (SBA-15, $D$ = 8.3 nm). Further details concerning the porous materials preparation and characterization are given in Section S1. The physical states of confined water corresponding to the two different hydration levels studied (75% and 33 % RH) are also sketched in Figure 1. Based on water sorption isotherms[35] and previous DSC studies,[36] (see also Section S2 for details on the hydration method), the porosity is completely filled with water for the larger RH value while water is adsorbed at the pore surface, leaving an empty pore center for the lower RH value.

The complex dielectric function $\varepsilon^*(f) = \varepsilon'(f) - i\varepsilon''(f)$, where $f$ is the frequency of the applied oscillating electrical field and $\varepsilon'$ and $\varepsilon''$ are the real and imaginary parts, respectively, was measured from 1 to $10^6$ Hz in the temperature range from 143 to 277 K for the different hydrated samples. Data acquisition and fitting procedures are detailed in Section S3.

**2.2. Capillary filled porous matrices at 75% RH.** The dielectric loss $\varepsilon''(\omega)$ of the different water-filled PMOs and MCM-41 porous matrices are presented for a selection of four temperatures in Figure 2. Isochronal representations of the dielectric loss as a function of the temperature ε″(T) are illustrated in Figure S1 and S2.

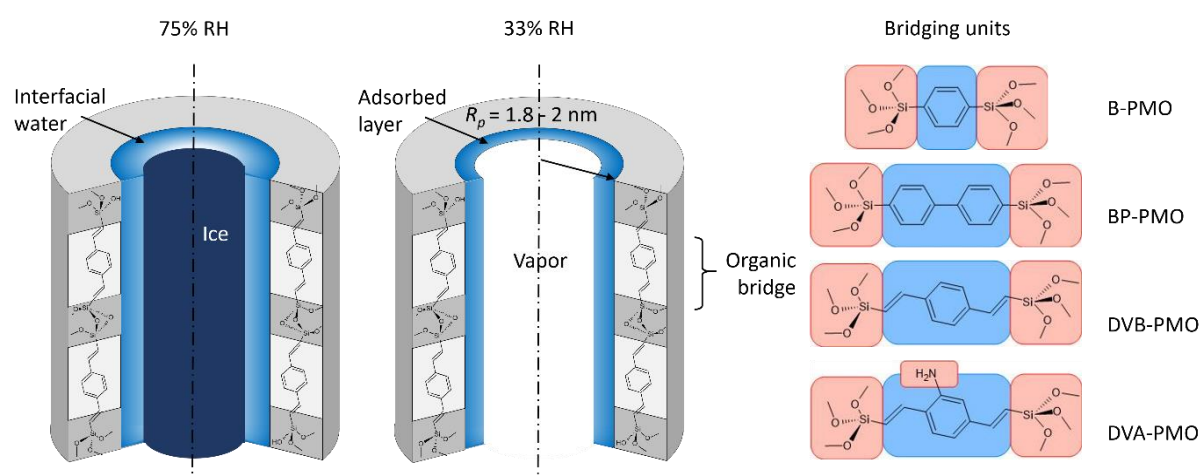

*Figure 1. Sketch of the different mesoporous materials and confined conditions studied. Left panel: sample filled at 75% RH, interfacial liquid coexists with ice. Middle panel: sample filled at 33% RH, water forms an adsorbed liquid layer. Right panel: different organic bridging units studied, with the hydrophilic and hydrophobic regions highlighted respectively by red and blue boxes.*

Two major processes were detected in these spectra, as indicated by arrows in Figure 2. Moreover, it should be mentioned three additional components, attributed to dc-conductivity, fast local water relaxation, and Maxwell-Wagner-Sillars polarization. The dc-conductivity is visible in Figure 2 at 203 K in the range of frequency below $10^2$ Hz. The fast local water relaxation emerged as a weak signal on the high frequency side of the relaxation peak at the lowest temperatures (c.a. below 155 K) as illustrated in Figure S3 for water-filled MCM-41 and discussed in ref. 28. Due to its small contribution to the dielectric function, over a limited temperature range, we refrained from performing a fit of this extra signal. Maxwell-Wagner-Sillars polarization appeared on the lowest frequency side of the spectra in the temperature range 195-210 K, and also in isochronal representations of the dielectric loss (Figures S1 and S2). This process was already observed in similar samples,[27,29] and related to the heterogeneous character of the porous media.[38,39]

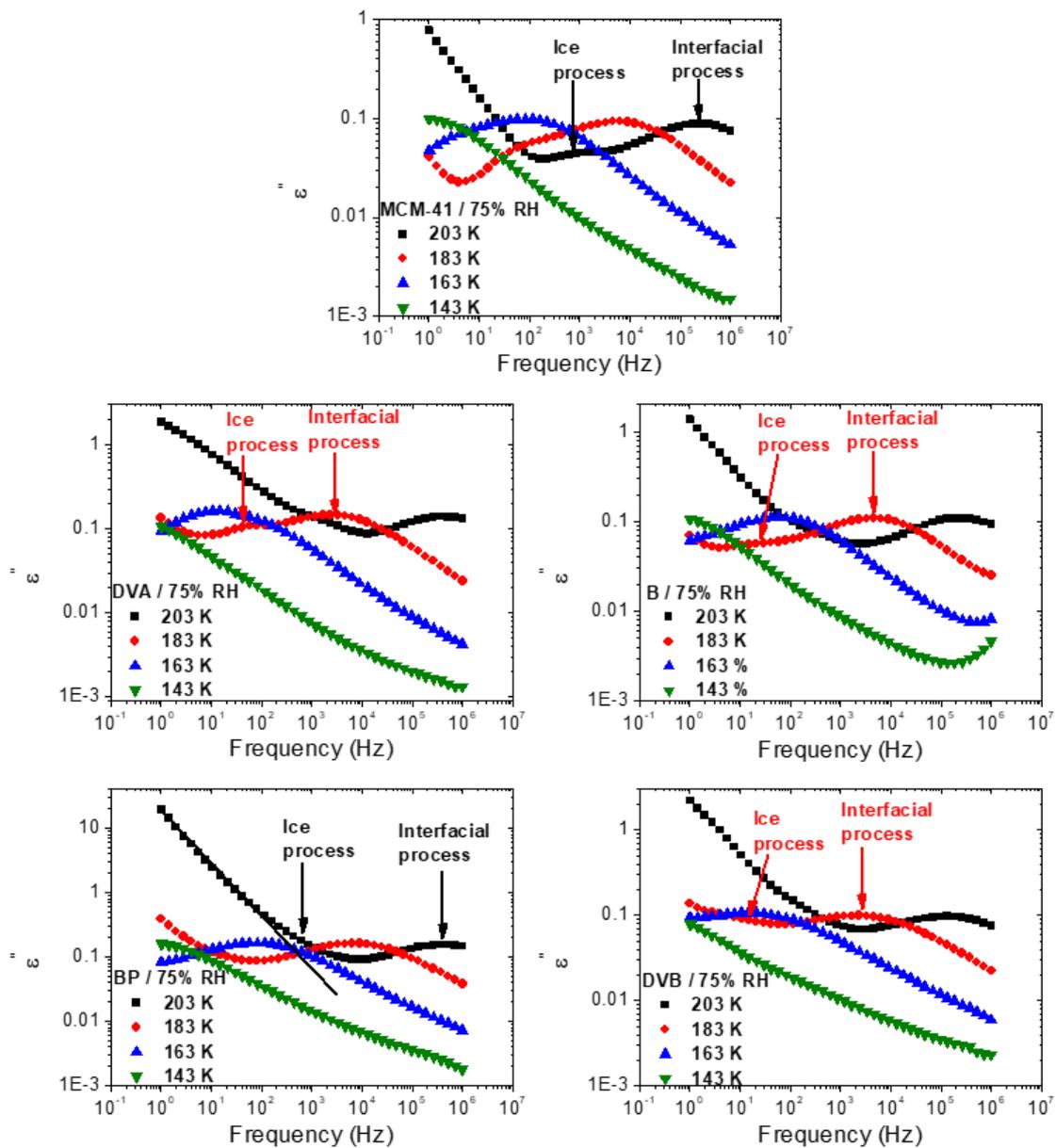

*Figure 2. Dielectric loss spectra ε"(ω) of water confined within the five matrices: MCM-41, DVA-PMO, B-PMO, BP-PMO, and DVB-PMO at different temperatures. Loading was carried out at 75% RH, ensuring the full loading of nanochannels by capillary condensation. Ice- and interfacial-processes are identified by arrows and the low frequency contribution is associated to MWS and/or conductivity.*

The two major relaxation processes can be related to the physical state of confined water. According to water physisorption isotherms, the entire porosity was filled after equilibrium with water vapor at 75% RH.[35,40] Moreover, the two relaxation processes entered the instrumental frequency range (1-$10^6$ Hz) for temperatures ranging from 143 to 223 K. This is below the freezing point of water confined in these matrices, that is in the range 227-237 K according to calorimetry.[36] In MCM-41, it has been shown that an interfacial layer of non-freezable water coexists with ice localized in the center of the pore.[27,37] This situation is likely to exist in PMOs too. Therefore, we attribute the two relaxation processes seen in water-filled PMOs to ice and interfacial water, in line with studies on porous silicas.[20,29,32] The slow process is associated to ice, and the faster one to interfacial water. Further support to this interpretation is provided by the present study from the effect of the pore size on the relative dielectric strength of both processes. This point is addressed in section S6 (cf. Figure S4 and S5), where water filled SBA-15 ($D$ = 8.3 nm) and MCM-41 ($D$ = 3.65 nm) are compared.

A model implying two Havriliak−Negami functions (HN functions) and dc-conductivity was fitted to the data. Very good fits were obtained for all the samples in the temperature range 143-223 K, as shown in Figure S4 and Figure S6 for MCM-41 and PMOs. The fitted HN parameters obtained are summarized in Table 1.

Table 1. HN parameters of the two processes obtained from the fit of the loss part of the dielectric function of water-filled MCM-41 and PMOs (75% RH) measured at T = 183 K.

| Process | HN Parameter | MCM-41 | DVA-PMO | B-PMO | BP-PMO | DVB-PMO |
|---|---|---|---|---|---|---|
| Ice | $\Delta\varepsilon$ | 0.0634 | 0.095 | 0.072 | 0.172 | 0.205 |
|  | $\alpha_{HN}$ | 0.8 | 0.9 | 0.8 | 0.6 | 0.6 |
|  | $\beta_{HN}$ | 1 | 1 | 1 | 1 | 1 |
| Interfacial | $\Delta\varepsilon$ | 0.438 | 0.636 | 0.5 | 0.742 | 0.501 |
|  | $\alpha_{HN}$ | 0.5 | 0.53 | 0.51 | 0.5 | 0.46 |
|  | $\beta_{HN}$ | 1 | 1 | 1 | 1 | 1 |
| $\Delta\varepsilon$(Interfacial)/$\Delta\varepsilon$(Ice) |  | 6.9 | 6.7 | 6.9 | 4.3 | 2.4 |

Where $\Delta\varepsilon$ is the dielectric strength, and $\alpha_{HN}$ and $\beta_{HN}$ ($0 < \alpha_{HN}$ ; $\alpha_{HN}\beta_{HN} \leq 1$) are fractional parameters describing, respectively, the symmetric and asymmetric broadening of the complex dielectric function, and characterize the relaxation heterogeneity[41].

The two HN exponents $\alpha_{HN}$ and $\beta_{HN}$ did not vary significantly with temperature, and were thereafter considered as temperature independent fitting parameters. Moreover, for the interfacial liquid water, they did not depend either on the type of porous material with $\alpha_{HN} \approx 0.5$ and $\beta_{HN} \approx 1$. The most noticeable feature concerns the value of the dielectric strength of the interfacial water process relative to that of the ice-process Δε(Interfacial)/Δε(Ice). Its value presented a systematic dependence on the nature of the matrix. The value of Δε(Interfacial)/Δε(Ice) averaged over the temperature range 173-203 K is illustrated in Figure 3. It was about 7 for MCM-41, DVA-PMO and B-PMO, and reduced to respectively 4 and 3 for BP-PMO and DVB-PMO. For SBA-15, the value is further decreased to 1.5, which may find a specific interpretation related to the much smaller surface-to-volume ratio of SBA-15 compared the other matrices. Notwithstanding the fact that absolute values of the dielectric permittivity are inaccessible due to the heterogeneous nature of the sample (cf. discussion in section S6) the large variation of the relative dielectric strength observed for PMOs suggests that the amount of water involved in the interfacial layer may depend, not only on the pore size, but on the pore surface chemistry, and the resulting interaction with water.

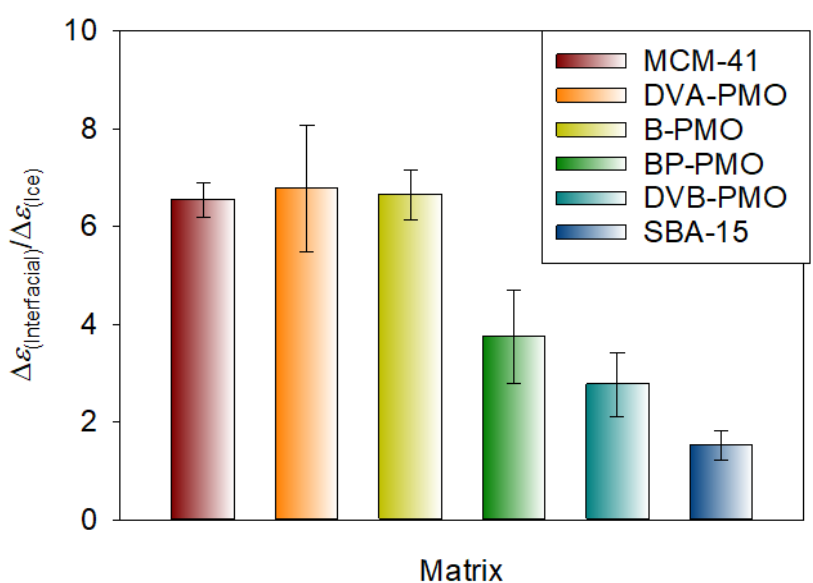

*Figure 3 Δε(Interfacial)/Δε(Ice) averaged over the temperature range 173-203 K for water confined in matrices loaded at 75% RH. Error bars illustrate the maximum temperature variations*

Two parameters controlling the water-surface interaction have been identified in previous studies of PMOs based on water physisorption[40] and multidimensional solid-state NMR spectroscopy studies.[35] The first is the length of the organic bridges that act as spacers between H-bonding silica units, as sketched in Figure 1. It was shown that decreasing the length of hydrophobic bridging units, which was the case when comparing divinylbenzene (DVB-PMO) and benzene (B-PMO), results in a more hydrophilic material.[40] The second parameter is the intrinsic hydrophilicity of the bridging unit, and especially its capability to form H-bonds. In the present study, this applies only for divinylaniline (DVA-PMO) that comprises an H-bonding amino group. This makes it more hydrophilic than divinylbenzene (DVB-PMO) and biphenyl (BP-PMO) bridging units that have comparable size but no polar group.[35] Based on these two parameters, Mietner et al. rationalized the relative hydrophilicity of series of PMOs matrices as quantified by the relative amount of adsorbed water below capillary condensation.[42] Interestingly, the systematic evolution of the ratio between the relative dielectric strengths of the interfacial water and the ice relaxation processes shown in Figure 3 follows the similar trend. It indicates that the fraction of water involved in the non-freezing interfacial layer is maximum for the more hydrophilic matrices (i.e. pure silica, PMOs with small hydrophobic bridging unit or long H-bonding bridging unit) and gradually decreases with the PMO hydrophilicity for large non-polar bridging units.

We now discuss the temperature dependence of reorientational dynamics of water confined in the five matrices (75% RH). The relaxation times assigned to the ice- and interfacial water processes, as well as the Maxwell-Wagner process when detected, are shown in Arrhenius representation in Figure 4. The two relaxation processes were grouped in grey areas in the relaxation map. For the two porous silicas (MCM-14 and SBA-15) our data were found in quantitative agreement with results from literature, as illustrated Figure S7.[27]

The dynamics of ice confined in MCM-41 and PMOs is only 2-3 orders of magnitude slower than the interfacial liquid in the temperature range analyzed, and faster than bulk ice. It is worth mentioning that the nature of the confined ice (hexagonal/cubic, stacking disordered) is still under ongoing discussion.[43] However, assuming that the dynamics of confined ice involves mechanisms that are similar to the bulk (i.e. orientational or ionic defects),[44] its higher mobility is likely attributable to significant structural distortion. Also, there is a large dispersion among the relaxation times of the ice-process in the different matrices. This is probably due to different levels of disorder in the ice. However, the relaxation times all retained an Arrhenius-like temperature dependence with a comparable activation energy $E_a$ which was in the range from 40 to 42 kJ.mol$^{-1}$ (± 4 kJ.mol$^{-1}$ for DVA-PMO, ± 0.5 kJ.mol$^{-1}$ for the other systems), suggesting that the underlying dynamic process in confined ice is barely affected by the nature of the matrix.

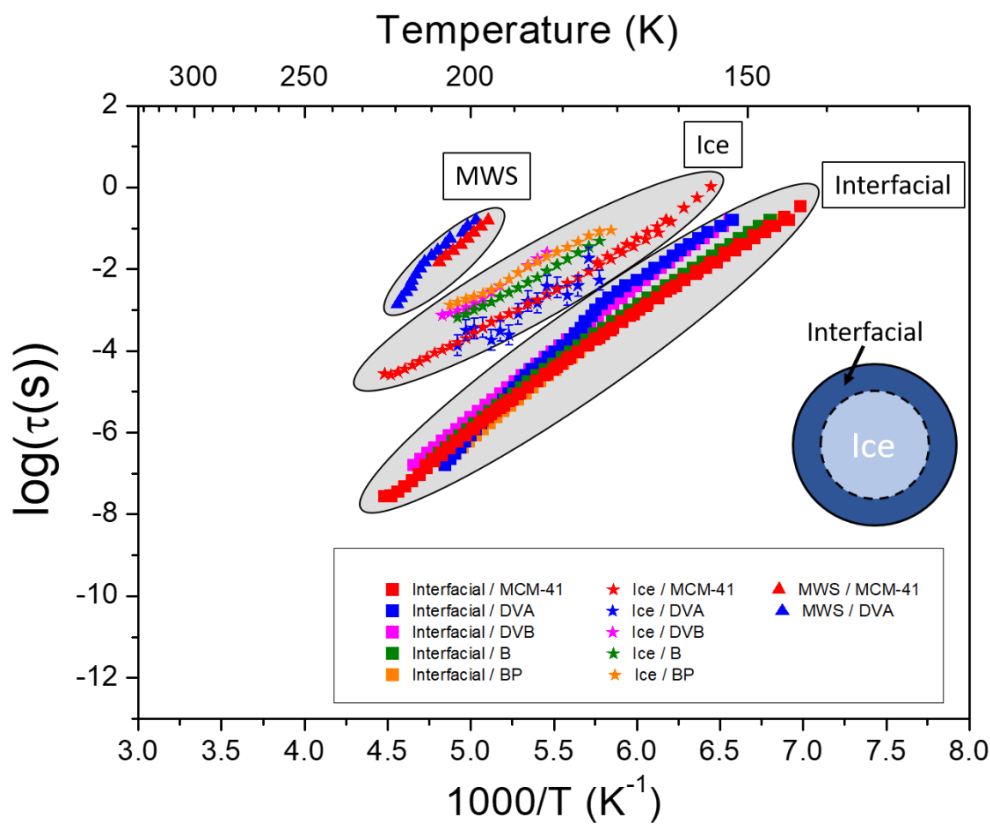

*Figure 4. Temperature dependence of relaxation times of the MWS-, ice- and interfacial-processes of water confined into the five matrices: MCM-41, DVA, B, BP and DVB. Loading was carried out at 75% RH, ensuring the complete filling of nanochannels by capillary condensation. Unless specified, errors bars are smaller than the symbol size.*

The second relaxation process, attributed to the reorientational dynamics of the interfacial water, was also found similar for the different confining matrices. It does not follow a unique Arrhenius dependence in the entire temperature range but displays a cusp around 180 K. For MCM-41, this observation has been discussed in terms of a crossover from a α-like relaxation above 180 K to a β-like process below 180 K, although this issue is still a matter of active debates that go beyond the scope of the present study.[7,14] However, we have adhered to the commonly held idea that the strictly Arrhenius behavior of deeply undercooled water is rather a characteristic feature of the lowest temperature region (i.e. below the crossover at 180 K). It was conjectured by Cerveny et al.[15] that this revealed the universal nature of the low-temperature dynamics of confined water. Therefore, Arrhenius fits of the data were carried out in the temperature region below the crossover at about 180 K, and the obtained parameters are presented in Table 2.

Table 2 Arrhenius parameters of the interfacial water relaxation time obtained by fitting the temperature region below 180K

| Porous matrix label | $\tau_\infty$ (s) | $E_a$ (kJ.mol$^{-1}$) |
|---|---|---|
| MCM-41 | ~ $10^{-22}$ | 50 ± 0.5 |
| DVA-PMO | ~ $10^{-18}$ | 50 ± 0.5 |
| B-PMO | ~ $10^{-19}$ | 51 ± 0.5 |
| BP-PMO | ~ $10^{-20}$ | 51 ± 0.5 |
| DVB-PMO | ~ $10^{-19}$ | 53 ± 0.5 |

The values of pre-exponential factor $\tau_\infty$ are very small ($10^{-18}$ – $10^{-22}$ s) compared to expected relaxation times that describe the high temperature limits of local orientation fluctuations in the liquid state (approximatively $10^{-13}$ – $10^{-14}$ s).[45] The similar observations of unphysical values of the pre-exponential factor $\tau_\infty$ of the dielectric relaxation of confined water have been reported by different groups.[27,29] This indicates that the molecular origin of the relaxation process attributed to the interfacial water might change at higher temperature. This phenomenon has been shown for water-filled MCM-41, which presents a crossover at 225 K as illustrated in Figure S7.[27]

Despite large variations in the pre-exponential factor, the values of the activation energy remained comparable, in the range from 50 to 53 ± 0.5 kJ.mol$^{-1}$ for all the five samples. This indicates that the surface chemistry has no significant influence on the relaxation process of the interfacial layer. This important finding extends previous conclusion made from $^2$H STE NMR experiments on water confined in surface modified MCM-41.[26] More generally, this support a number of studies relating to confined water in different pore sizes [29,31], and to water mixtures.[46,47] Despite very different conditions, they all converge to the existence of a same relaxation process that appears for temperatures below a dynamical crossover located at around 180 K. As such, this relaxation process could be indicative of the universal character of the corresponding dynamics of water.[46]

In light of this interpretation, one may conclude that the dynamics of water in the liquid interfacial region is governed by water-water interactions, and less sensitive to water-pore surface interactions. In fact, the value of the activation energy corresponds to the energy required for breaking at least two hydrogen bonds.[26] In the interfacial phase, a water molecule can interact with either another interfacial water molecule or with a water molecule from the ice phase. In this case, the water-pore surface interaction seems to play a secondary role, as demonstrated by the negligible influence of the matrix surface chemistry.

Different observations were made in a recent NMR study of $D_2O$ confined in SBA-15 comprising surfaces functionalized with different amino acids.[48] The value of the correlation times for the reorientation dynamics was found to increase systematically with the acidic character of the residue, while, as in the present study, the corresponding activation energies barely changed. It should be stressed that the study was performed in the temperature range above 200K, i.e. above the crossover usually seen in confined water.[28] Also, the surface treatment was achieved by chemical grafting of relatively long chains inducing softness in the confining medium, while the organic bridging units of PMOs are incorporated into the matrix walls. From the valuable comparison of both works, it emerges that the degree of coupling between the dynamics of interfacial water and the pore surface must depend on several parameters, in particular the softness of the confinement, the flexibility, polarity and ionic charge of the surface functional groups, and the temperature range. A different situation is also awaited when water forms an adsorbed monolayer.

**2.3 Adsorbed water on the pore surface of PMOs (33% RH).** The dynamics of water adsorbed on the surface of MCM-41 and PMOs was studied for samples loaded at a reduced relative pressure (33% RH). According to the water physisorption isotherms,[35,40] this condition ensures being situated before the capillary condensation filling. Thus, only a layer of adsorbed water can be formed on the pore wall, leaving the pore center empty.

The dielectric loss $\varepsilon''(\omega)$ of water adsorbed at 33% RH on the surface of the different matrices is illustrated in Figure 5 for a selection of 4 temperatures. Isochronal representations of $\varepsilon''(T)$ are also presented in Figure S8. Compared to the completely filled matrices, the low-frequency dc-conductivity was reduced, which is easily conceivable due to restricted translation mobility of adsorbed molecules. A low frequency process attributed to Maxwell-Wagner polarization was also observed. Most importantly, water-filled DVA-PMO, additionally exhibited two relaxation processes, while only one relaxation process was observed for all the other systems. The corresponding dielectric losses were fitted by respectively two and one HN functions (cf. Figure S9). It is worth noting that the shape parameters $\alpha_{HN}$ and $\beta_{HN}$ are smaller for the partially filled matrices compared to the completely filled matrices (Figure S6 and Table 1). It can be related to a broader distribution relaxation times, arising from dynamic inhomogeneity and different environments.[49] The reduced value of $\beta_{HN}$ for the partially filled samples (except DVA) could also indicate the emergence of a secondary high-frequency process. However, this hypothesis is not confirmed by masterplot curves, Figure S10, that are rather consistent with the presence of only one relaxation process. The resulting relaxation times are shown in Figure 6. As shown in Figure S11, present results fully agree with previous findings for partially water-filled silica pores.[27,29]

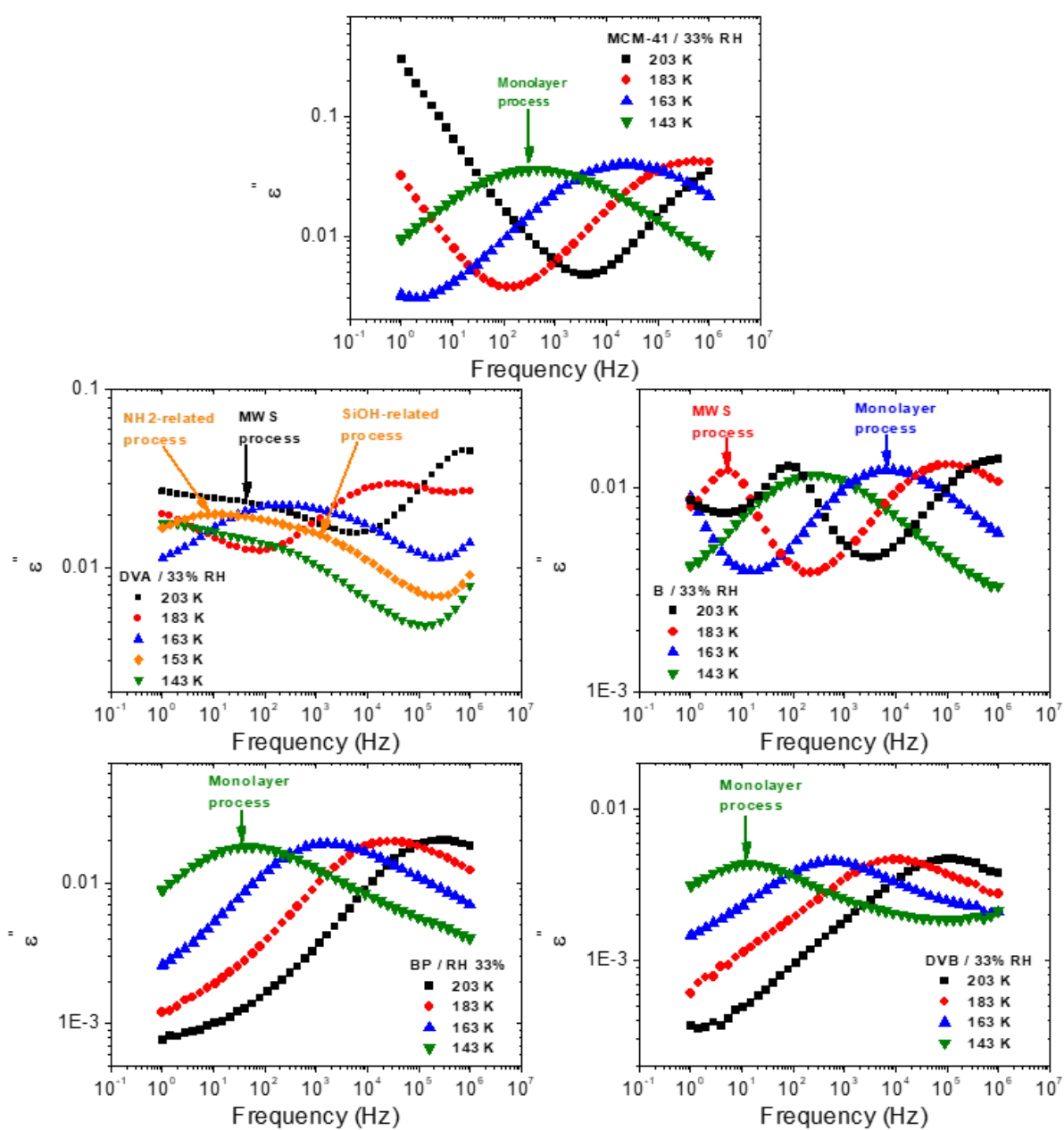

*Figure 5. Dielectric loss spectra ε''(ω) of water confined within the five matrices: MCM-41, DVA-PMO, B-PMO, BP-PMO, and DVB-PMO at different temperatures. Loading was carried out at 33% RH, below the capillarity-condensation, resulting in monolayer of water at the inner surface and empty center. MWS- and monolayer-processes are identified by arrows.*

We first discuss the systems showing a single relaxation mode, *i.e.* MCM-41, B-PMO, BP-PMO, and DVB-PMO. The relaxation mode of each system appeared on a comparable timescale and exhibited a qualitatively similar temperature dependence. This suggests that they have a common origin. This unique relaxation process is consistent with the presence of a single phase formed by the surface layer adsorbed at the pore wall.[29] In line with the nomenclature often used in literature,[27,29] the process arising from the adsorbed surface layer is later designated as the monolayer process. This does not imply that the microscopic picture of water forming a single molecular layer that covers uniformly the pore surface is strictly valid. Indeed, it cannot be ruled out the existence of water clusters on the silica surface, a situation which would be also likely for PMOs due to the modulated surface chemistry.

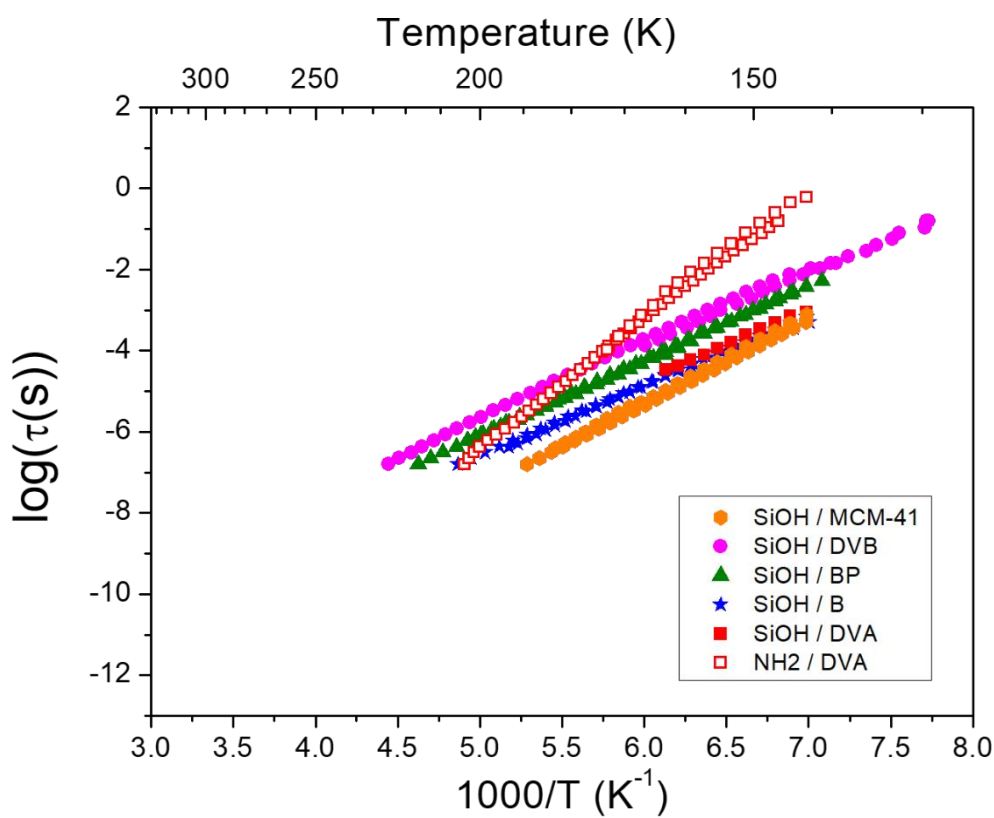

*Figure 6. Temperature dependence of relaxation times of water adsorbed on the pore surface of the five matrices: MCM-41, DVB-PMO, BP-PMO, B-PMO, and DVA-PMO. Loading was carried out at 33% RH, below the capillarity-condensation, resulting in monolayer of water at the inner surface and empty center. Unless specified, errors bars are smaller than the symbol size.*

Interestingly, this monolayer process cannot be confused with the interfacial water process seen in the samples loaded at 75% RH. First, it is about 2-3 decades faster than the interfacial water process, as also illustrated in Figure S10.[27,29] Moreover, the relaxation time of the monolayer processes presents an Arrhenian temperature dependence in the entire temperature range analyzed. The Arrhenius fitting parameters are summarized in Table 3. Importantly, this analysis reveals that the nature of the water-surface interactions, which is controlled by the pore chemistry of PMOs, differently affects the monolayer and interfacial water processes. Indeed, the activation energy is reduced to values in the range from 33 to 35 ± 0.5 kJ.mol$^{-1}$ for PMOs compared to 40 ± 0.5 kJ.mol$^{-1}$ for MCM-41.

We infer that the activation energy reflects the different environments around the adsorption sites found in MCM-41 and PMOs. Due to their polarity and their ability to form H-bonds, silanols are the preferential adsorption sites for water molecules. This was confirmed by quantum mechanical calculations for a model silica surface by Saengsawang *et al.*[50] Among different possible configurations, this study has shown that the strongest binding energy was achieved for cluster formed by a water interacting with two silanols. This configuration is also likely to occur on the pore surface of MCM-41 silica matrices. However, it must be rarer for PMOs due to the larger average distance between silanols. More precisely, considering that the organic and silica bridging units are alternately occupying periodic positions along the pore z-axis, it might still be possible for a water molecule to be bounded to two silanols that occupy radially adjacent positions.

Table 3 Arrhenius parameters of monolayer processes for confined water in the five matrices at 33% RH.

| Porous matrix label | $\tau_\infty$ (s) | $E_a$ (kJ.mol$^{-1}$) |
|---|---|---|
| MCM-41 | ~ 10$^{-18}$ | 40 ± 0.5 |
| DVA-PMO[a] | ~ 10$^{-23}$ | 57 ± 0.5 |
| DVA-PMO[b] | ~ 10$^{-16}$ | 33 ± 0.5 |
| B-PMO | ~ 10$^{-16}$ | 33 ± 0.5 |
| BP-PMO | ~ 10$^{-16}$ | 35 ± 0.5 |
| DVB-PMO | ~ 10$^{-15}$ | 34 ± 0.5 |

[a]Process related to water interacting with the amine groups.

[b]Process related to water interacting with the silanol groups.

However, this is impossible for two axially adjacent silanols because they are spatially separated by an organic bridging unit. Indeed, we recall that the repetition distance of silica units along the z-axis (cf. Figure 1) is $\Delta z$ = 0.76 nm for B-PMO and $\Delta z$ = 1.2 nm for BP-PMO and DVB-PMO, which exceeds the size of a single water molecule. In PMOs, single water-silanol bound is still possible, but energetically weaker,[50] which we designate as the possible origin of the smaller activation energy for the dipolar relaxation mode. This interpretation also rationalizes the fact that, despite different repetition distance, a similar relaxation process was observed for the three PMOs.

We finally consider the specific case of water adsorbed in DVA-PMO, which presented two distinct relaxation processes. It is obvious from Figure 6 that the timescale of the fastest process agrees well with that of the single relaxation process discussed above for the three other PMOs. This is also true for the value of the activation energy, which equals 33 kJ.mol$^{-1}$. We therefore attribute this process to the rotational motion of the water molecules adsorbed on the silanols of the silica unit. The activation energy of the second relaxation process (57 kJ.mol$^{-1}$) is almost twice as large as the first one, and also larger than the monolayer process for MCM-41.

It is worth noting that the coexistence of two relaxation peaks was also reported in a previous study for MCM-41 with comparable pore size ($D$ = 3.6 nm).[27] This situation was obtained for intermediate levels of hydration, and the two relaxation processes were attributed to coexisting adsorbed water monolayer and capillary condensed water. In this study, the samples were hydrated by exposing them to 100% RH, but the hydration step was limited to a water load below complete filling. The level of hydration *H* was expressed in terms of water mass uptake per mass of dry sample, which should not be confused with RH. Only a monolayer relaxation was observed at the hydration level (*H* = 8%), while the second relaxation emerged at *H* = 15%. Although the hydration method used in that study complicates the estimation of the actual RH, it is very likely that the hydration level (*H* = 15%), where two relaxation processes were reported for MCM-41, was indeed located within the hysteresis loop of the capillary condensation. This could explain the coexistence of capillary filled pores with other pore surfaces only covered with a monolayer.

In the present study, the two values of the relative humidity (33% and 75% RH) were selected to be properly situated outside the hysteresis loop of the capillary condensation, that is to say, in two regions of the water adsorption isotherm where both the adsorption and desorption branches were superimposed. This excludes the possible coexistence of unfilled and capillary filled regions, and also the hypothetical formation of metastable states. In fact, based on the water physisorption isotherms of PMOs (cf. supporting information in ref. [35]), we could estimate that the hydration level for DVA-PMO (33% RH) was *H* = 7%, which is comparable to *H* = 8% in the above-mentioned study, where only a single monolayer process was observed for MCM-41.[27]

Therefore, the case of DVA-PMO requires another interpretation. In order to attribute this additional process, two useful remarks can be done: (i) compared to the other PMOs, and especially DVB-PMO, the only specificity of DVA-PMO arises from the presence of an amine H-bonding group, (ii) the activation energy of this second process is approaching the value obtained for the interfacial water in the case of capillary filled samples (i.e. 75% RH). According to (i), it seems natural to associate this process to the presence of a secondary adsorption site located on the bridging unit. In fact, the existence of two different adsorption sites in DVA-PMO has been demonstrated for capillary filled samples by 2D $^1$H-$^{29}$Si and $^1$H-$^{13}$C HETCOR NMR.[35] In these experiments, two sets of NMR cross peaks revealed that water closely interacted with the organic unit, and with the silica part of the pore surface of DVA-PMO. In contrast, for a non H-bonded organic bridging unit, *i.e.* BP-PMO, the cross peaks were restricted to water-silica interactions. This supports the hypothesis that the second relaxation process actually arises from water molecules interacting with the amine sites. Although interesting, this picture alone does not provide a clear explanation for the large value of the activation energy (ii). In fact, nitrogen being less electronegative than oxygen, the strength of the H-bond associated to this second adsorption site is expectedly weaker than that associated to silanol. In our opinion, the most sensible interpretation cannot only involve individual adsorbed molecule, but requires the formation of clusters of water molecules on the pore surface before capillary condensation. This phenomenon can obviously be favored by the spatial modulation of the surface interaction, induced by the specific surface chemistry of PMO. The higher conformational flexibility of the organic bridging units with respect to surface silanols, may also confer a stabilizing effect on the H-bonding amino groups of the DVA moieties. In contrast to adsorbed monolayer, water clusters involve both water-surface and water-water interactions, which is consistent with the larger value of the activation energy of the corresponding relaxation process.

## 3. CONCLUSIONS

We performed a systematic DRS study on the reorientation dynamics of water confined in mesoporous silica with different pore diameters (SBA-15, MCM-41) and four organosilicas (PMOs) with different surface chemistries and surface interactions. Two values of the filling fractions were achieved by controlling the value of the relative humidity maintained during the hydration process. The first condition (75% RH) stands above the completion of the capillary condensation, which ensures a full loading of the porous volume with water and no bulk excess. The second (33% RH) is located below the onset of capillary condensation, which implies that water molecules are adsorbed on the inner surface of the channels leaving the center of the pores empty.

For completely filled pores, partial crystallization occurred in the temperature range studied. Two relaxation processes were attributed respectively to ice and to unfreezable interfacial water that is sandwiched between the pore surface and the surface of distorted ice, in line with previous findings.[20,27–29] Geometrical considerations based on the relative variation of the dielectric strength of these modes as a function of the pore size of the two silicas ($D$ = 8.3 and 3.6 nm) were found consistent with the estimated thickness of the interfacial layer ($e \approx 0.6$ nm). This value corresponding to about two molecular sizes agrees with the thickness derived from thermoporometric studies.[37]

For capillary filled PMOs (75% RH) with comparable pore sizes but different surface chemistry, a systematic evolution of the dielectric strength of the interfacial layer relative to ice was observed. A simplified description, based on the ratio of the relative dielectric strengths of the two relaxation processes, indicated a correlation between the amount of water involved in the interfacial layer and the chemistry of the pore surface. As a general trend, the contribution from the interfacial layer was found decreasing with increasing the surface hydrophobicity. Nevertheless, there was found no significant influence of the surface chemistry on the water reorientation dynamics in the interfacial layer, in terms of relaxation time and activation energy. This indicates that the dynamics of water molecules forming the interfacial layer between ice and the pore surface is governed by water-water interactions, and that the surface chemistry plays a secondary role.

This finding contrasts with the case of water molecules adsorbed on the pore surface at low partial pressure (33% RH). In this case, depending on the surface chemistry, we have identified three distinct situations. (i) For MCM-41, a single Arrhenius-like relaxation process was observed. The value of its relative dielectric strength was about half that of the interfacial process measured at full loading, which is consistent with the formation of an adsorbed monolayer. Its faster dynamics and its smaller activation energy (40 kJ.mol$^{-1}$) further demonstrate that the nature of the monolayer is obviously different from the interfacial water discussed above. In MCM-41, rather than water-water interaction, the rotational dynamics of monolayer molecules is determined by the H-bond interaction with the surface silanols. (ii) For PMOs, a similar relaxation process was also observed, but its activation energy was reduced to 33-35 kJ.mol$^{-1}$ compared to 40 kJ.mol$^{-1}$ for MCM-41. While H-bonding silanols are still present on the surface of PMOs, they are distributed periodically along the pore axis. We have shown that the repetition distance, determined by the length of the organic bridging unit, hinders the formation of multiple H-bonds with adsorbed water molecules. As a result, we infer that the smaller energy required to break the water-surface binding interaction consequently reduces the activation energy of the rotational dynamics of adsorbed water. (iii) For DVA-PMO, an additional relaxation process with Arrhenius-like temperature dependence was observed. It was attributed to the specific capability of DVA to act as a second H-bonding adsorption sites for surface water molecules, due to the presence of an amine group. The relatively large value of the activation energy (57 kJ.mol$^{-1}$) of this second process points to the formation of clusters of H-bonded water molecules in the vicinity of the DVA units. The stabilization of water clusters could benefit from the higher conformational flexibility of DVA organics with respect to silanols. At present, this interpretation still needs to be supported by additional sets of data, and would greatly profit from a microscopic picture as offered by molecular simulation. Interestingly, this picture definitively contrasts with the simplified view of adsorbed water molecules forming a uniformly distributed monolayer on the pore surface, which was suggested for MCM-41.

This highlights the new possibility of controlling the properties of confined liquids thanks to the spatial modulation of the surface interaction, induced by the specific surface chemistry of PMOs, both with regard to the equilibrium and non-equilibrium, transport behavior.[51] In particular the exploration of capillarity-driven flow phenomena,[52,53] where the surface modulation affects the resulting Laplace pressures and effective viscosities and thus hydraulic permeabilities could be rewarding in the future.

- ASSOCIATED CONTENT

**Supporting Information** Further information on Materials (Section S1, Table S1); sample hydration method (Section S2); dielectric spectroscopy method (Section S3.); Dielectric loss of water confined in MCM-41 and PMOs loaded at 75% RH (Figure S1, Figure S2); secondary weak process of water filled MCM-41 (Figure S3); pore size effects on water filled MCM-41 and SBA-15 silicas (Figure S4, Figure S5, Table S2, Table S3); dielectric fitted functions for water filled PMOs (Figure S6); temperature dependence of the different relaxation times for water filled MCM-41 and SBA-15 (Figure S7); dielectric loss of water adsorbed in PMOs at 33% RH (Figure S8); dielectric fitted functions for water adsorbed in PMOs at 33% RH (Figure S9); Masterplot curves of the monolayer process in the five matrices (Figure S10); temperature dependence of the different relaxation times for water adsorbed in mesoporous MCM-41 silicas at two RH (Figure S11); Highlighting crystallization and melting processes in SBA-15 loaded at 75% RH (Figure S12); references (Section S14).

- ACKNOWLEDGEMENTS

This work was conducted in the frame of the DFG-ANR collaborative project (Project NanoLiquids No. ANR-18-CE92-0011-01, DFG Grant No. FR 1372/25-1- Project number 407319385, and DFG Grant No. Hu850/11-1- Project number 407319385), which is acknowledged. Support was received from Rennes Metropole and Europe (European Regional Development Fund - CPER PRINT2TAN). We thank Dr. Malina Bilo for providing the B-PMO sample. It is a pleasure to acknowledge Prof. Andreas Schönhals for fruitful discussions and critical reading of the manuscript. We also acknowledge the scientific exchange and support of the Center for Molecular Water Science (CMWS).

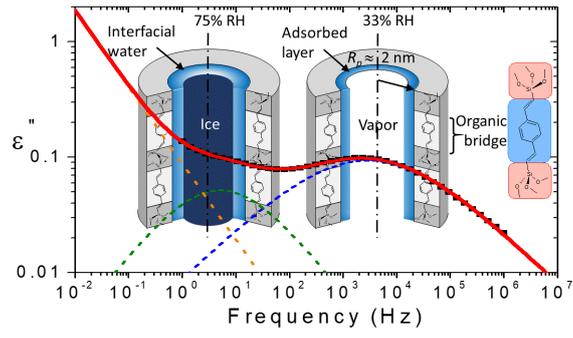

Toc Graphic

# Supporting Information for

# Influence of Pore Surface Chemistry on the Rotational Dynamics of Nanoconfined Water

Benjamin Malfait[1], Aicha Jani[1], Jakob Benedikt Mietner[2], Ronan Lefort[1], Patrick Huber[3,4,5,*], Michael Fröba[3,*], and Denis Morineau[1,*]

[1]Institute of Physics of Rennes, CNRS-University of Rennes 1, UMR 6251, F-35042 Rennes, France

[2]Institute of Inorganic and Applied Chemistry, University of Hamburg, 20146 Hamburg, Germany

[3]Institute for Materials and X-ray Physics, Hamburg University of Technology, 21073 Hamburg, Germany

[4]Centre for X-Ray and Nano Science CXNS, Deutsches Elektronen-Synchrotron DESY, 22603 Hamburg, Germany

[5]Centre for Hybrid Nanostructures CHyN, Hamburg University, 22607 Hamburg, Germany

*Email:patrick.huber@tuhh.de

*Email:froeba@chemie.uni-hamburg.de

*Email:denis.morineau@univ-rennes1.fr

Table of content

## S1. Materials

Periodic Mesoporous Organosilicas (PMOs) powders were prepared according to the following procedure. NaOH and the alkyltrimethylammonium bromide surfactant were dissolved in deionized water. The bis-silylated precursors of the form $(EtO)_3Si-R-Si(OEt)_3$ (R = organic unit) were added at room temperature, and the mixtures were stirred for 20 hours. The mixtures were transferred into a Teflon-lined steel autoclave and statically heated to 95 °C or 100 °C for 24 h. The resultant precipitate was collected by filtration and washed with 200 ml deionized water. After drying at 60 °C, the powder was extracted with a mixture of ethanol and hydrochloric acid (EtOH:HCl (37 %), 97:3, v/v) using a Soxhlet extractor. The porosity and the pore structure of the dried materials were characterized by powder X-ray diffraction and nitrogen physiosorption.[1] In this work, we used four different PMOs materials with different organic bridging groups: divinylbenzene (DVB), divinylaniline (DVA), benzene (B) and biphenyl (BP). By studying these PMOs, we can study both the effect of hydrophobicity (DVA-PMO vs DVB-PMO) and repetition distances of silica/organic units along the pore axis that is related to the length of the organic unit (B-PMO vs BP-PMO), as illustrated in Table S1.

The mesoporous materials MCM-41 silicas were prepared according to a procedure similar to that described elsewhere[2] and already used in previous works.[3–5] Hexadecyl ammonium bromide was used as template to get a mesostructured 2D-hexagonal array of aligned channels with pore diameter $D = 3.65$ nm, as confirmed by nitrogen adsorption, transmission electron microscopy and neutron diffraction. The SBA-15 mesoporous silicas were prepared using a procedure described elsewhere,[6,7] with slight modifications of the thermal treatments to optimize the final structure of the product.[8] Nonionic triblock copolymer (Pluronic $P_{123}$): $(EO)_{20}(PO)_{70}(EO)_{20}$ was used as a template to obtain a mesostructured 2D-hexagonal array of aligned channels with a pore diameter $D = 8.3$ nm, and porous volume $V_P = 1.0$ cm$^3$ g$^{-1}$.

The structural parameters of all the matrices are summarized in the Table S1.

*Table S1. Structural and chemical parameters of the mesoporous matrices. The hydrophilic and hydrophobic regions are highlighted respectively by red and blue boxes.*

| Matrix label | Organic bridging unit | Chemical formula | Repetition distance (nm)[a] | Pore surface ($m^2 \cdot g^{-1}$)[b] | Pore volume ($cm^3 \cdot g^{-1}$)[b] | Pore diameter (nm)[b] |
|---|---|---|---|---|---|---|
| MCM-41 | … | 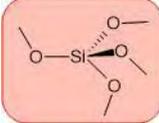 | … | 1077 | 0.89 | 3.65 |
| SBA-15 | … | 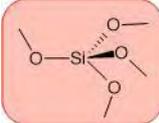 | … | 880 | 1.0 | 8.3 |
| B-PMO | Benzene | 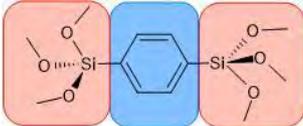 | 0.75 | 815 | 0.68 | 4.1 |
| BP-PMO | Biphenyl | 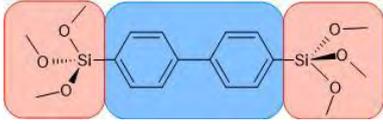 | 1.20 | 786 | 0.445 | 3.5 |
| DVB-PMO | Divinyl-benzene | 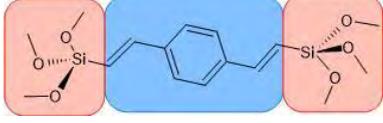 | 1.18 | 864 | 0.99 | 4.1 |
| DVA-PMO | Divinyl-aniline | 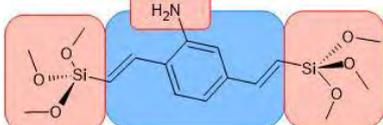 | 1.18 | 1223 | 0.89 | 3.5 |

[a]Evaluated from the (00l) Bragg reflections.

[b]Evaluated from the nitrogen physisorption isotherms.

## S2. Sample hydration

To prepare hydrated matrices, it was found preferable to impose the relative pressure rather than the mass fraction of water that fills the porous materials. It ensures that all the different water-filled materials result from the equilibrium with water vapor at the same chemical potential, which makes their comparison more reliable from the thermodynamic point of view. The filling procedure that was applied in this study can be considered as an experimental realization of the grand canonical thermodynamic ensemble, where the chemical potential of water molecules in the gas phase is imposed by the relative humidity (i.e. RH = $P_{water}/P_{sat}$). At a given temperature, the relation between the amount of water that fills each porous medium and the relative pressure of the saturating atmosphere is determined by the water-vapor adsorption isotherm. They were measured for series of PMOs that were similar to the PMOs used in the present study.[9] They were all characteristic of type IV isotherms: at low relative pressure, the isotherm exhibits a low adsorption region, followed by a pore capillary condensation step at an intermediate relative pressure (0.5-0.6, depending on the PMO), and reaches a plateau where the amount of the water increases only slightly with the increase of pressure.

Based on this principle, the matrices powders were compacted between two stainless steel electrodes to form a pellet, and then placed in a desiccator in the presence of a beaker containing a saturated aqueous solution of $MgCl_2$ and NaCl, and equilibrated at around 20 °C for 24 h. The resulting relative humidities (RH) were 33% for the $MgCl_2$ and 75% for the NaCl. The former RH is located below the onset of the capillary condensation, which results in water being adsorbed at the pore surface with an empty pore center. The second RH is above the capillary condensation, leading to the complete filling of the porosity with no excess bulk-like liquid.[9]

## S3. Dielectric spectroscopy method

The complex dielectric function $\varepsilon^*(f) = \varepsilon'(f) - i\varepsilon''(f)$, where $f$ is the frequency of the applied oscillating electrical field and $\varepsilon'$ and $\varepsilon''$ are the real and imaginary parts, respectively, was measured from 1 to $10^6$ Hz using a Novocontrol high-resolution ALPHA analyzer (Montabaur, Germany). Samples were prepared in parallel plate geometry between two stainless steel electrodes with a diameter of 20 mm and typical sample thickness $e$ of about 0.2 mm. The cryostat was pre-cooled in order to quench the sample rapidly to the starting temperature, $T$ = 277 K, ensuring a controlled nitrogen atmosphere that prevent any water adsorption/desorption once the sample is placed in the cryostat. The stability of the sample loading was also confirmed by comparing the spectra obtained during the cooling and heating branches.

Dielectric spectra were collected isothermally from 277 to 143 K in decreasing steps of 2 K. The same procedure was performed once 143 K was reached, but in increasing steps to 277 K. The sample temperature was regulated using a Quatro cryo-system nitrogen flux device that keeps temperature fluctuations within 0.5 K.

The spectra were analyzed in the isothermal representation of the dielectric loss ($\varepsilon''$ versus frequency at constant temperature $\varepsilon''(f; T = const)$). The contribution from conductivity was fitted to Eq. 1

$$\varepsilon''(f) = \frac{\sigma}{\varepsilon_0 2\pi f} \qquad (1)$$

where $\sigma$ is the dc conductivity, $\varepsilon_0$ is the vacuum permittivity. A sum of Havriliak-Negami (HN) functions was fitted to the experimental relaxation processes. Each HN function write as

$$\varepsilon^*(\omega) = \varepsilon_\infty + \frac{\Delta\varepsilon}{[1+(i\omega\tau_{HN})^{\alpha_{HN}}]^{\beta_{HN}}} \qquad (2)$$

with $\omega = 2\pi f$ and where $\Delta\varepsilon$ is the dielectric strength, $\tau_{HN}$ is the characteristic HN relaxation time, and $\alpha_{HN}$ and $\beta_{HN}$ ($0 < \alpha_{HN}; \alpha_{HN}\beta_{HN} \leq 1$) are fractional parameters describing, respectively, the symmetric and asymmetric broadening of the complex dielectric function. From the fitting parameters $\tau_{HN}$, $\alpha_{HN}$ and $\beta_{HN}$, a model-independent relaxation time $\tau$ associated to the maximum in the loss part of the complex dielectric function was determined according to: [10,11]

$$\tau = \tau_{HN} \left[ \frac{\sin\left(\frac{\alpha_{HN}\beta_{HN}\pi}{2+2\beta_{HN}}\right)}{\sin\left(\frac{\alpha_{HN}\pi}{2+2\beta_{HN}}\right)} \right]^{1/\alpha_{HN}} \qquad (3)$$

Isochronal representation of the dielectric loss ($\varepsilon''$ versus temperature at constant frequency $\varepsilon''(T; f = const)$) was also fitted with a superposition of k Gaussians functions[12] for relaxation processes and an exponential function for the conductivity.

When applicable, the temperature dependence of the relaxation times were described by the Arrhenius equation: [13]

$$\tau = \tau_\infty \times exp\left(\frac{E_a}{RT}\right) \qquad (4)$$

where $E_a$ is the apparent activation energy and $R$ is the ideal gas constant.

S4. Isochronal representation ε''(T) of the dielectric loss of water confined in MCM-41 and in the four PMOs loaded at 75% RH

Figure S1 and Figure S2 represent the dielectric loss ε''(T) for water confined in MCM-41 and in PMOs loaded at 75% RH, *i.e.* above the capillary-condensation, for three values of the frequency 1, 10 and 100 Hz. The physical state distribution in the pore is [Interfacial + Ice]. This representation allows to evidence a process at high temperature, which is attributed to the Maxwell-Wagner-Sillars process. The jump in the ε''(T) signal, which is frequency-independent, located at around -40 °C is associated to the melting of the ice phase.

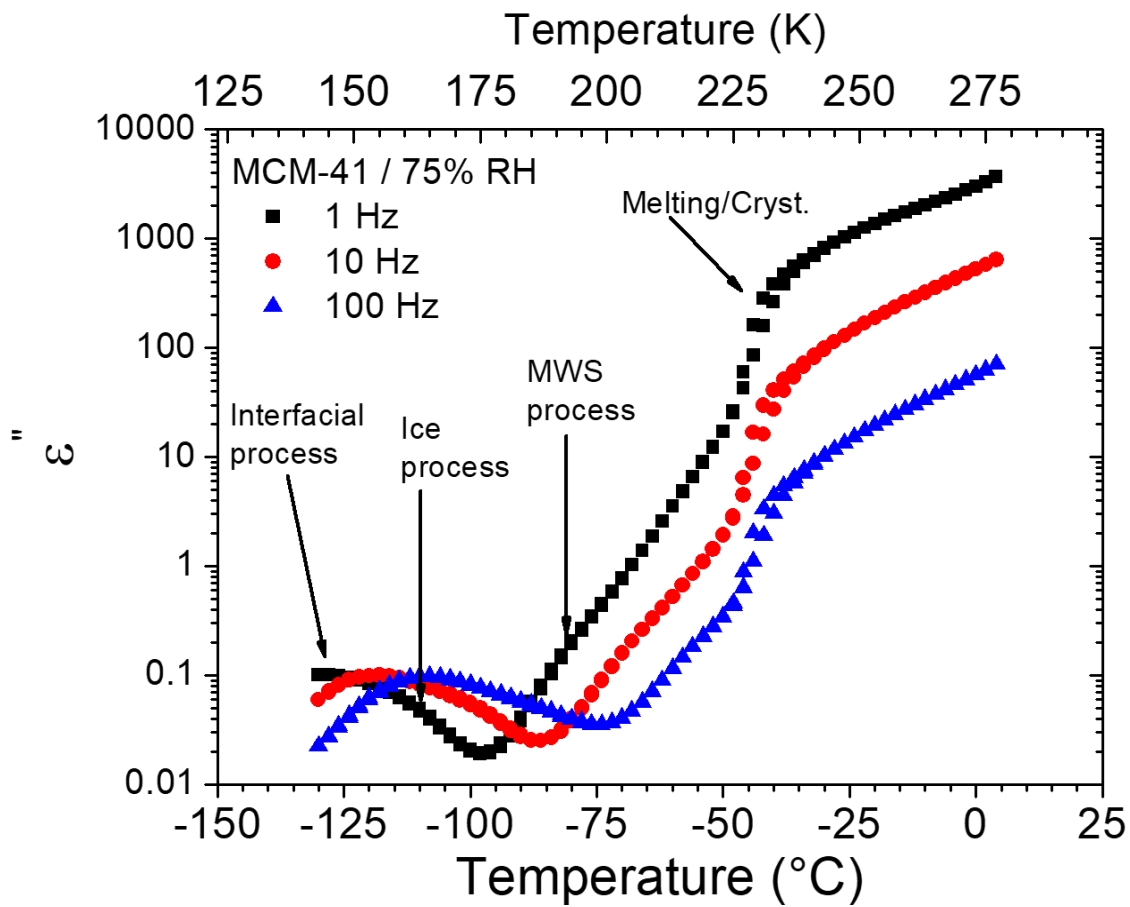

*Figure S1 Dielectric loss ε''(T) for water confined in MCM-41 loaded at 75% RH for three values of the frequency 1, 10 and 100 Hz.*

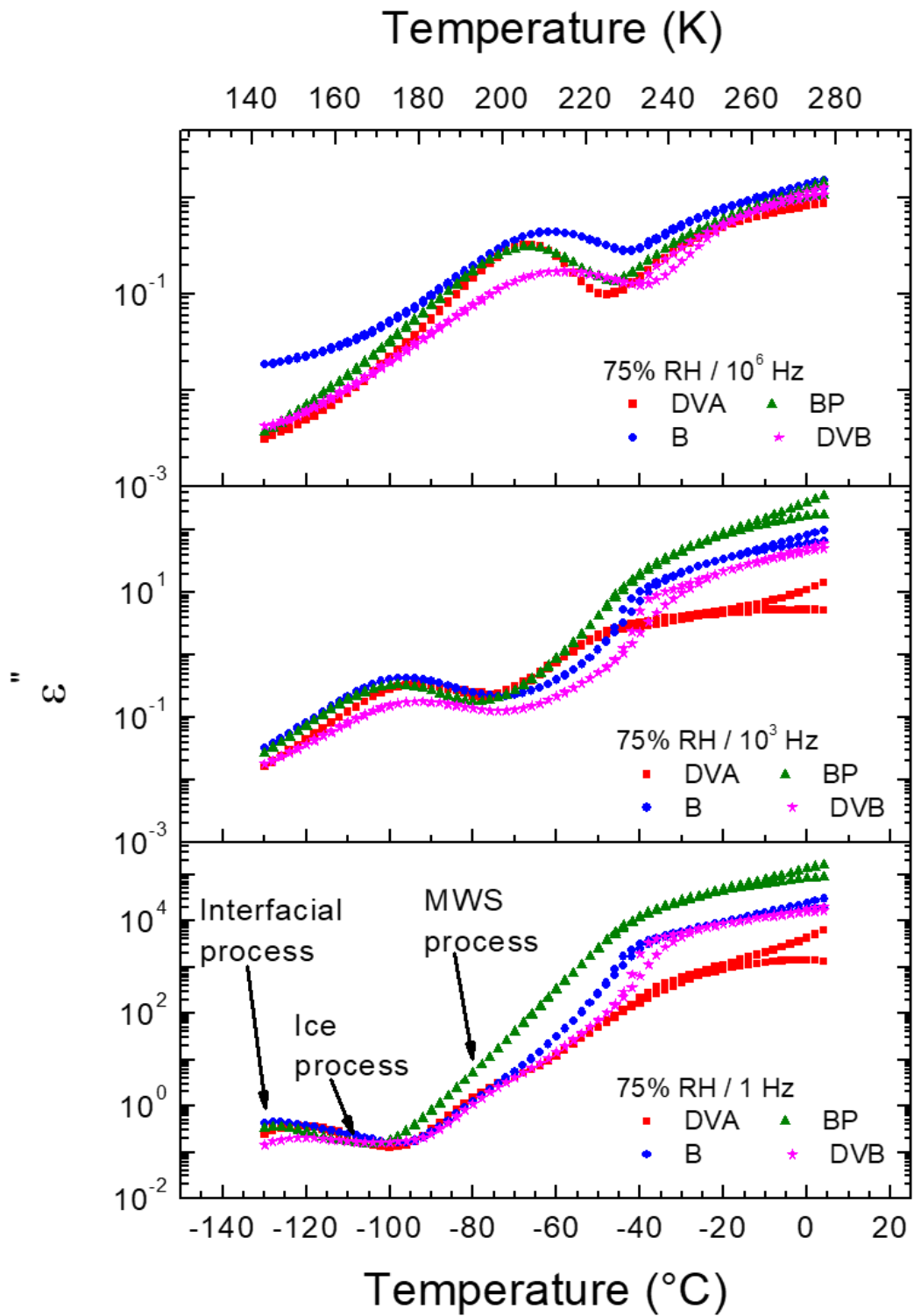

*Figure S2 Isochronal representation ε"(T) of the dielectric loss of water confined in the four PMOs loaded at 75% RH, for three values of the frequency 1, $10^3$ and $10^6$ Hz.*

## S5.     Secondary weak process of water filled MCM-41

Figure S3 represents the dielectric loss ε″(ω) versus frequency for water confined in MCM-41 loaded at 75% RH, *i.e.* above the capillary-condensation, at 153 K. The physical state distribution in the pore is [Interfacial + Ice]. A clear break in the slope of the ε″(ω) signal is observed in the high frequency side of the relaxation peak. In a log/log representation, HN functions present linear evolution against frequency. Therefore, a slope breaking highlights an excess process.

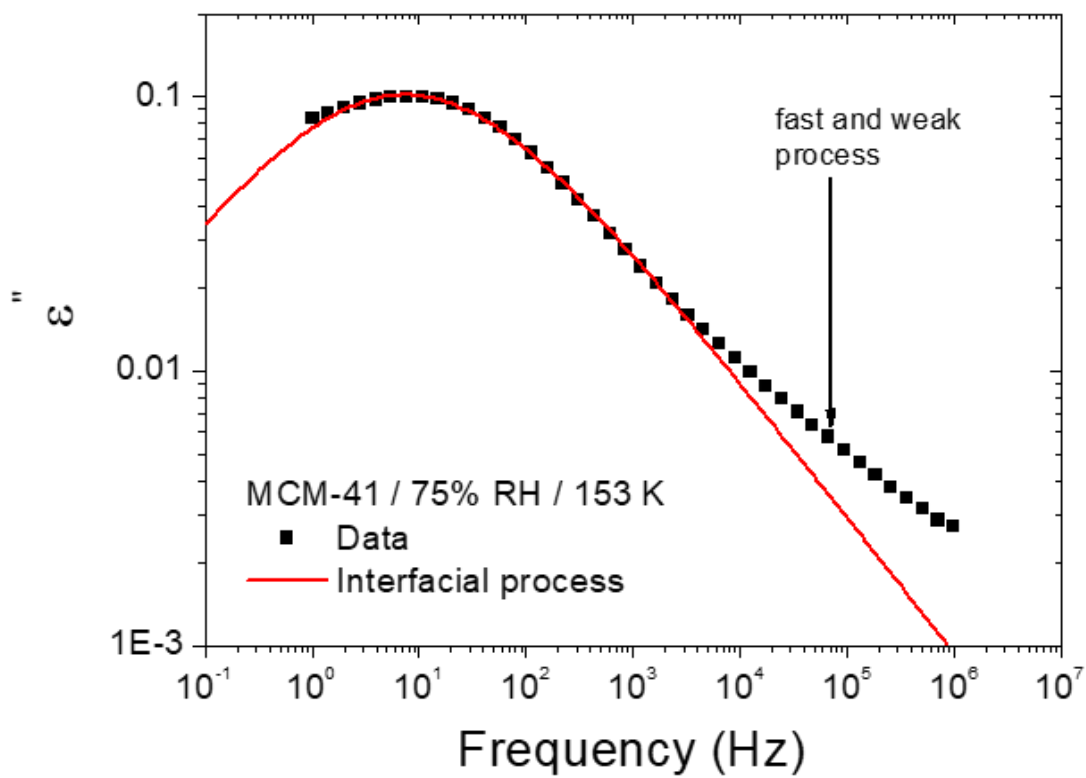

*Figure S3 Dielectric loss ε″(ω) versus frequency for water confined in MCM-41 loaded at 75% RH at 153 K.*

## S6. Pore size effects on water filled MCM-41 and SBA-15 silicas.

The dielectric loss ε″ measured at full loading (75% RH) and $T$ = 183 K for water filled SBA-15 and MCM-41 are illustrated in Figure S4.

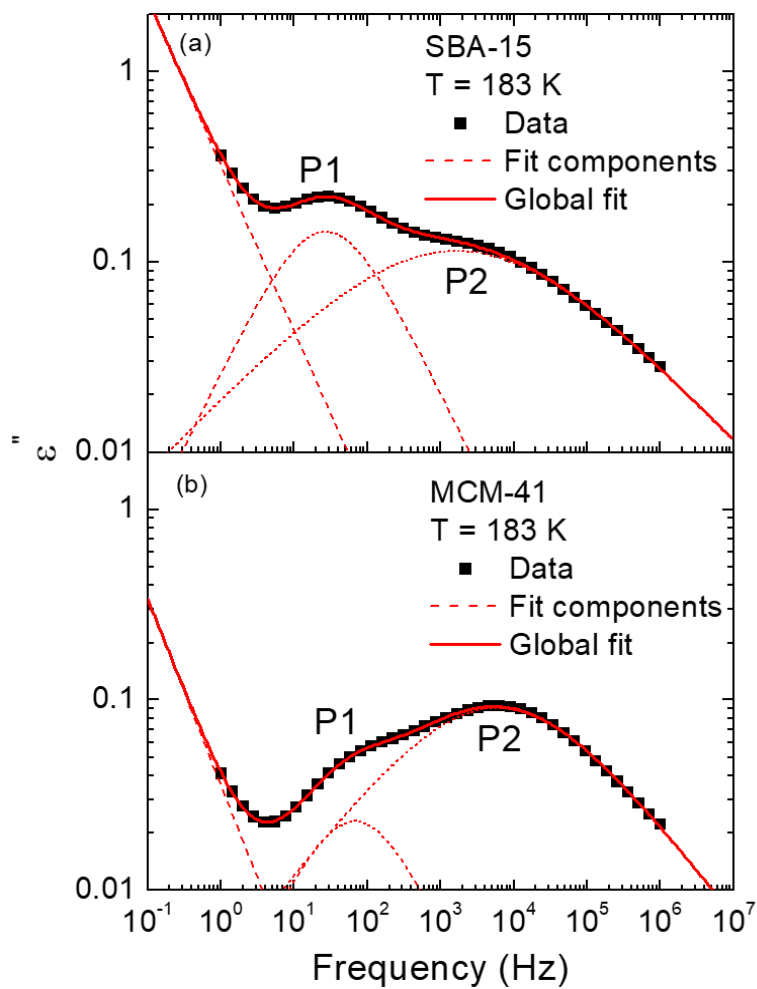

*Figure S4. Dielectric loss spectra ε″ of water confined into (a) SBA-15 and (b) MCM-41 at 183K, loaded at 75% RH. The solid lines are the global fit considering the sum of individual HN functions for each individual relaxation process while the low frequency component is fitted with a conductivity function. HN and conductivity functions are dashed lines.*

The fitting parameters obtained for the two relaxation processes P1 and P2 are provided in Table S2 for $T$ = 183 K.

Table S2. HN parameters of the two processes P1 and P2 obtained from the fit of the loss part of the dielectric function of water-filled MCM-41 and SBA-15 (75% RH) measured at $T$ = 183 K.

| Process | HN Parameter | MCM-41 | SBA-15 |
|---|---|---|---|
| P1 (Ice) | $\Delta\varepsilon$ | 0.0634 | 0.393 |
|  | $\alpha_{HN}$ | 0.8 | 0.8 |
|  | $\beta_{HN}$ | 1 | 1 |
| P2 (Interfacial) | $\Delta\varepsilon$ | 0.438 | 0.708 |
|  | $\alpha_{HN}$ | 0.5 | 0.4 |
|  | $\beta_{HN}$ | 1 | 1 |

At 183 K, the physical state of confined water in these matrices consists in two co-existing phases, ice and interfacial liquid water. Ice is localized in the center of the pore while the interfacial water is sandwiched between ice and matrix inner surface.[14,15]

We have derived a consistent image supporting this interpretation, based on the values of the dielectric strength and on geometric considerations. The sample measured is heterogeneous. It is a powder, which consists of porous material silica, confined water and intergranular voids. Due to the complexity of the resulting equivalent circuit, the determination of an absolute value of the dielectric strength is illusive. In order to reduce the impact of the sample-specific scaling factors related to different pellet thicknesses or powder packing fractions, the relative dielectric strength of each process, i.e. $\Delta\epsilon(P1)/\Delta\epsilon(P2)$ was compared for the two confining matrices by the ratio $f_{\Delta\epsilon}$ defined as

$$f_{\Delta\epsilon} = \frac{\Delta\epsilon(P1_{MCM})/\Delta\epsilon(P2_{MCM})}{\Delta\epsilon(P1_{SBA})/\Delta\epsilon(P2_{SBA})} \qquad (5)$$

Concomitantly, a second ratio $f_V$ was defined based on geometric considerations. It assumed that the interfacial layer has a typical thickness $e \approx 0.6$ nm for both confined systems, based on the evaluation made in the literature by thermoporometry,[14,16] and illustrated schematically in Figure S5. The ratio $f_V$ expresses as

$$f_V = \frac{V(Ice_{MCM})/V(Interfacial_{MCM})}{V(Ice_{SBA})/V(Interfacial_{SBA})} = \frac{(R_{p,MCM}-e)^2 / [R_{p,MCM}^2 - (R_{p,MCM}-e)^2]}{(R_{p,SBA}-e)^2 / [R_{p,SBA}^2 - (R_{p,SBA}-e)^2]} \qquad (6)$$

where $V(Ice_{MCM})$, $V(Ice_{SBA})$, $V(Interfacial_{MCM})$ and $V(Interfacial_{SBA})$ are the volumes occupied by, respectively, the core ice in MCM-41, the core ice in SBA-15, the interfacial water in MCM-41 and the interfacial water in SBA-15. $R_{p,MCM}$ and $R_{p,SBA}$ are respectively the pore radius of MCM-41 and SBA-15.

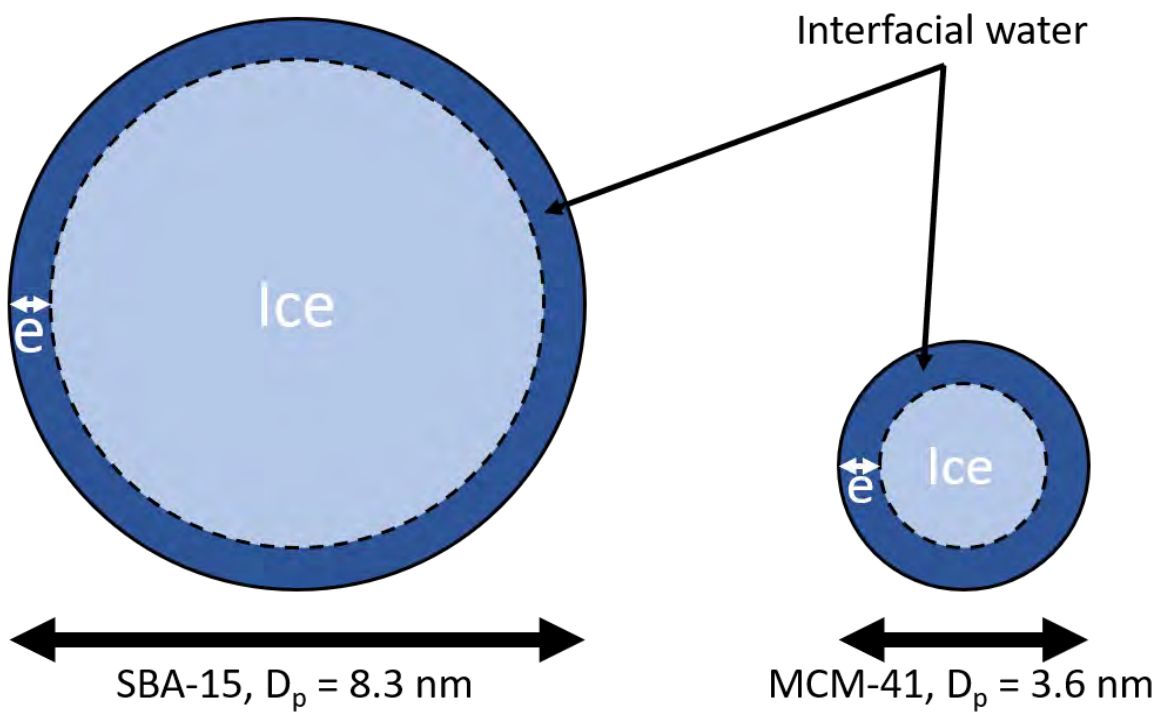

*Figure S5. Schematic view of the physical state distribution in a radial section for confined water in SBA-15 (left) and MCM-41 (right) at T = 143 K. The dimensions scales are preserved between the two matrices. The interfacial liquid layer has a thickness of 0.6 nm. $D_p$ is the diameter of the pore.*

It is worth noting that the main approximations made in this calculation: (i) the dielectric strength in Eq. 5 was supposed to be proportional to the quantity of water present in each region (ii) the geometry in Eq. 6, was assumed perfectly cylindrical with a fixed interlayer thickness conforming cryoporometry studies,[14,16] and (iii) the microporosity in SBA-15 was not considered.

The values of $f_{\Delta\epsilon}$ obtained in the temperature range from 163 to 203 K were distributed from 0.19 to 0.26 (i.e. $f_{\Delta\epsilon}$= 0.224 +/- 15%) as shown in Table S3 They are in fair agreement with the ratio evaluated from geometrical consideration, $f_V$=0.293, given the simplicity of the description and the number of underlying assumptions. This further confirms the attribution of processes P1 and P2 to ice and interfacial water, respectively. This description of the relative strength of the confined ice process had not been done at a quantitative level so far, although it was already reported as gaining intensity with increasing hydration level by Gainaru *et al*.[17] for water in connective tissue proteins collagen and elastin. It should be also noted a non-monotonous temperature variation of $f_{\Delta\epsilon}$, although this phenomenon presently lacks firm interpretation. More specifically, we could not conclude on the hypothetical temperature dependence of the interfacial layer thickness *e*, as expected for interfacial melting,[18] since the temperature range studied is well below the melting/freezing hysteresis loop as determined by DSC.[19]

**-70 °C**

| | | MCM-41 | SBA-15 |
|---|---|---|---|
| P1 / Ice | Δε | 0.0675 | 0.496 |
| | α$_{HN}$ | 0.8 | 0.8 |
| | β$_{HN}$ | 1 | 1 |
| P2 / Interfacial | Δε | 0.42 | 0.577 |
| | α$_{HN}$ | 0.5 | 0.4 |
| | β$_{HN}$ | 1 | 1 |

$$\frac{\Delta\varepsilon(Ice_{MCM})/\Delta\varepsilon(Interfacial_{MCM})}{\Delta\varepsilon(Ice_{SBA})/\Delta\varepsilon(Interfacial_{SBA})} = 0.187$$

**-80 °C**

| | | MCM-41 | SBA-15 |
|---|---|---|---|
| P1 / Ice | Δε | 0.0648 | 0.464 |
| | α$_{HN}$ | 0.8 | 0.8 |
| | β$_{HN}$ | 1 | 1 |
| P2 / Interfacial | Δε | 0.427 | 0.628 |
| | α$_{HN}$ | 0.5 | 0.4 |
| | β$_{HN}$ | 1 | 1 |

$$\frac{\Delta\varepsilon(Ice_{MCM})/\Delta\varepsilon(Interfacial_{MCM})}{\Delta\varepsilon(Ice_{SBA})/\Delta\varepsilon(Interfacial_{SBA})} = 0.205$$

**-90 °C**

| | | MCM-41 | SBA-15 |
|---|---|---|---|
| P1 / Ice | Δε | 0.0634 | 0.393 |
| | α$_{HN}$ | 0.8 | 0.8 |
| | β$_{HN}$ | 1 | 1 |
| P2 / Interfacial | Δε | 0.438 | 0.708 |
| | α$_{HN}$ | 0.5 | 0.4 |
| | β$_{HN}$ | 1 | 1 |

$$\frac{\Delta\varepsilon(Ice_{MCM})/\Delta\varepsilon(Interfacial_{MCM})}{\Delta\varepsilon(Ice_{SBA})/\Delta\varepsilon(Interfacial_{SBA})} = 0.261$$

**-100 °C**

| | | MCM-41 | SBA-15 |
|---|---|---|---|
| P1 / Ice | Δε | 0.0683 | 0.406 |
| | α$_{HN}$ | 0.8 | 0.8 |
| | β$_{HN}$ | 1 | 1 |
| P2 / Interfacial | Δε | 0.444 | 0.686 |
| | α$_{HN}$ | 0.5 | 0.4 |
| | β$_{HN}$ | 1 | 1 |

$$\frac{\Delta\varepsilon(Ice_{MCM})/\Delta\varepsilon(Interfacial_{MCM})}{\Delta\varepsilon(Ice_{SBA})/\Delta\varepsilon(Interfacial_{SBA})} = 0.260$$

**-110 °C**

| | | MCM-41 | SBA-15 |
|---|---|---|---|
| P1 / Ice | Δε | 0.0497 | 0.371 |
| | α$_{HN}$ | 0.8 | 0.8 |
| | β$_{HN}$ | 1 | 1 |
| P2 / Interfacial | Δε | 0.464 | 0.711 |
| | α$_{HN}$ | 0.5 | 0.4 |
| | β$_{HN}$ | 1 | 1 |

$$\frac{\Delta\varepsilon(Ice_{MCM})/\Delta\varepsilon(Interfacial_{MCM})}{\Delta\varepsilon(Ice_{SBA})/\Delta\varepsilon(Interfacial_{SBA})} = 0.205$$

Table S3. HN parameters obtained for the two processes P1 (Ice) and P2 (Interfacial) at several temperature for confined water in MCM-41 loaded at 75% RH.

## S7. Dielectric fitted functions for water filled PMOs

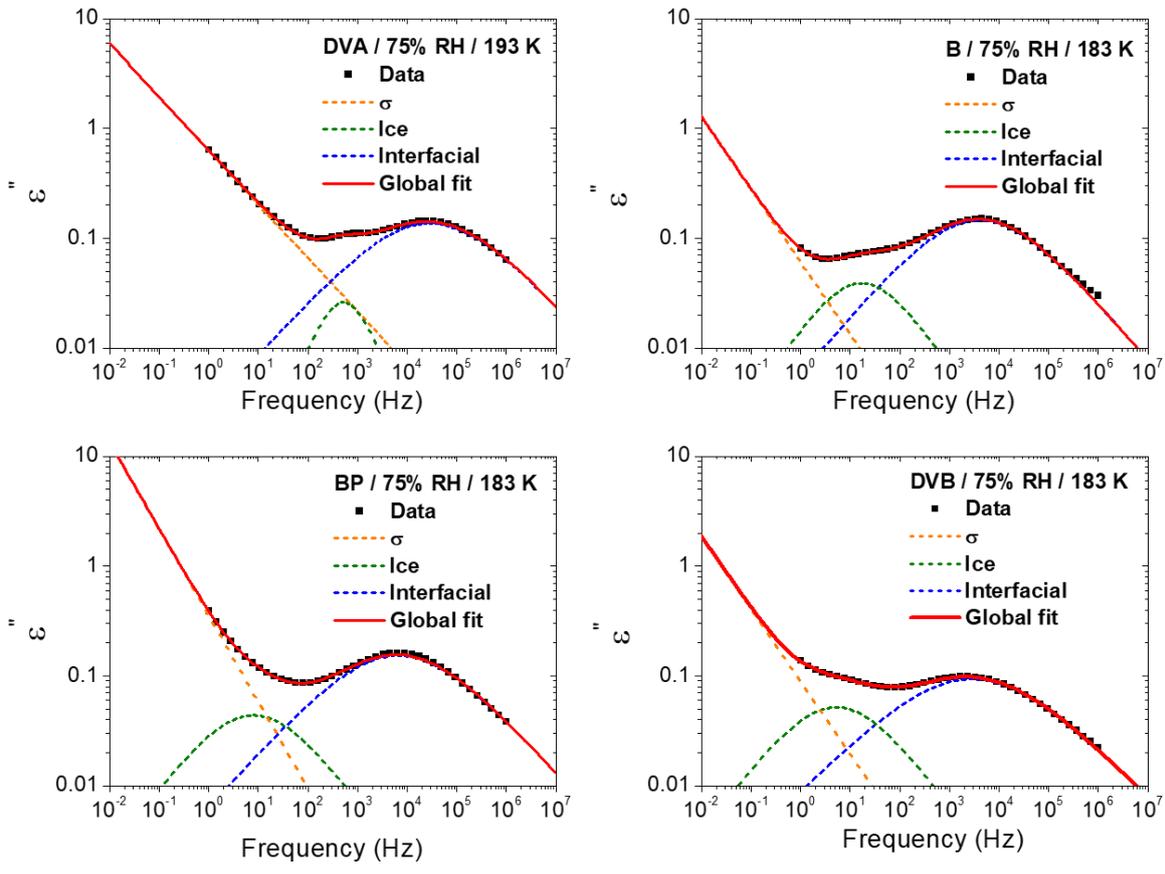

*Figure S6 Dielectric functions fitted to the experimental loss of water-filled PMOs loaded at 75% RH.*

## S8. Temperature dependence of the different relaxation times for water filled MCM-41 and SBA-15.

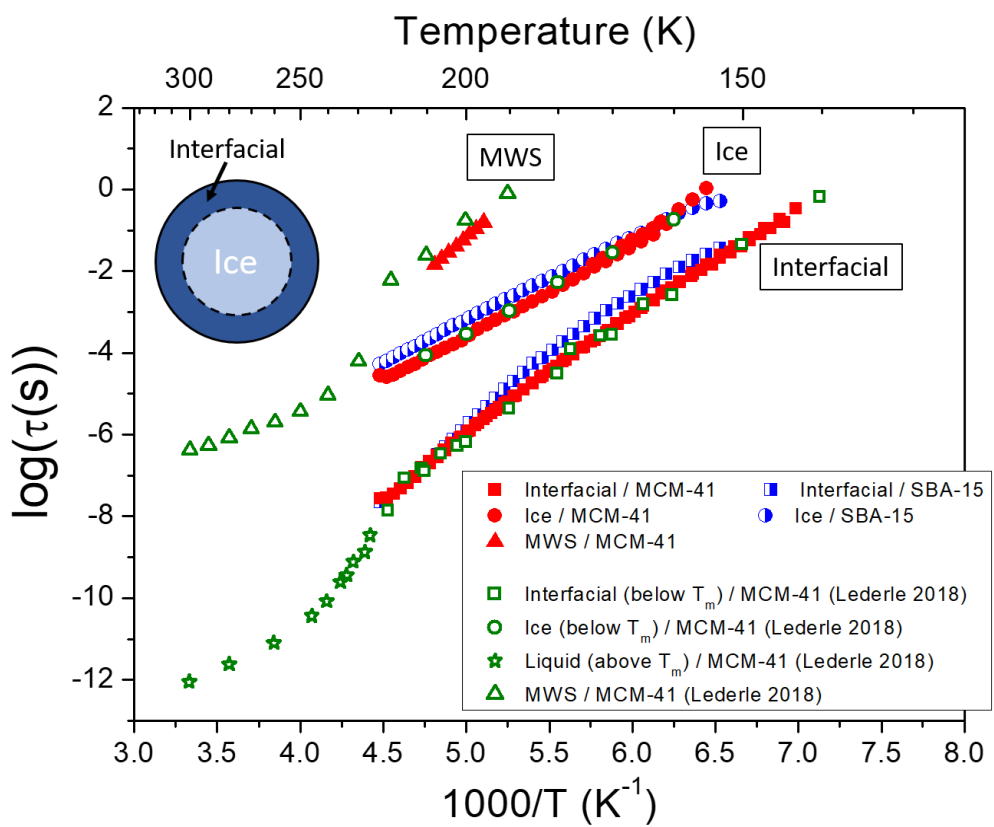

*Figure S7. Temperature dependence of relaxation times of the MWS, ice- and interfacial-processes of water confined into MCM-41. Loading was carried out at 75% RH, ensuring the full loading of nanochannels by capillary condensation. Additionally, data from literature[15] were plotted in order to strengthen the identification of processes.*

S9. Isochronal representation ε″(T) of the dielectric loss of water adsorbed in PMOs at 33% RH

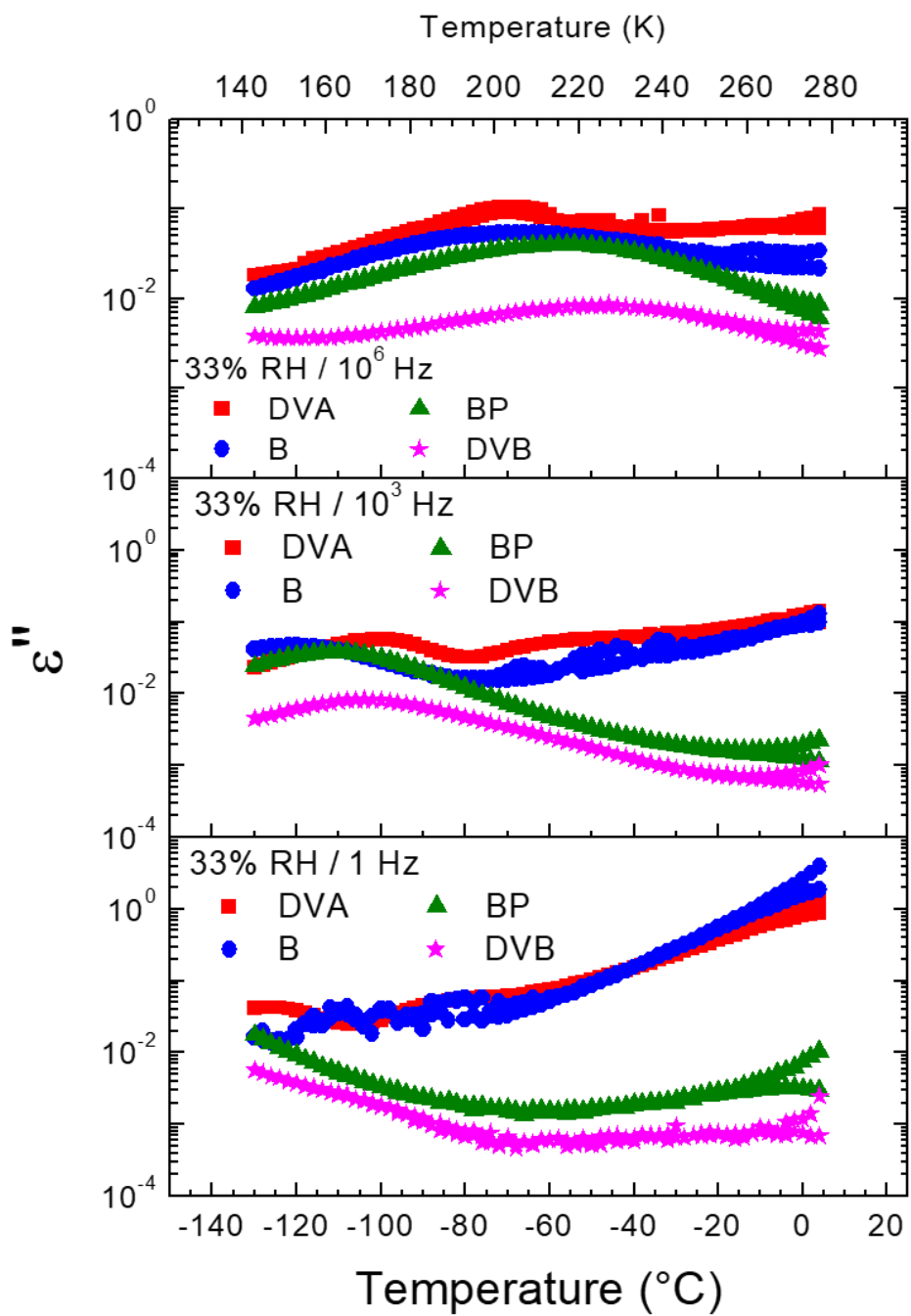

*Figure S8 Isochronal representation ε″(T) of the dielectric loss of water adsorbed in the four PMOs at 33% RH, for three values of the frequency 1, $10^3$ and $10^6$ Hz.*

## S10. Dielectric fitted functions for water confined in PMOs loaded at 33% RH

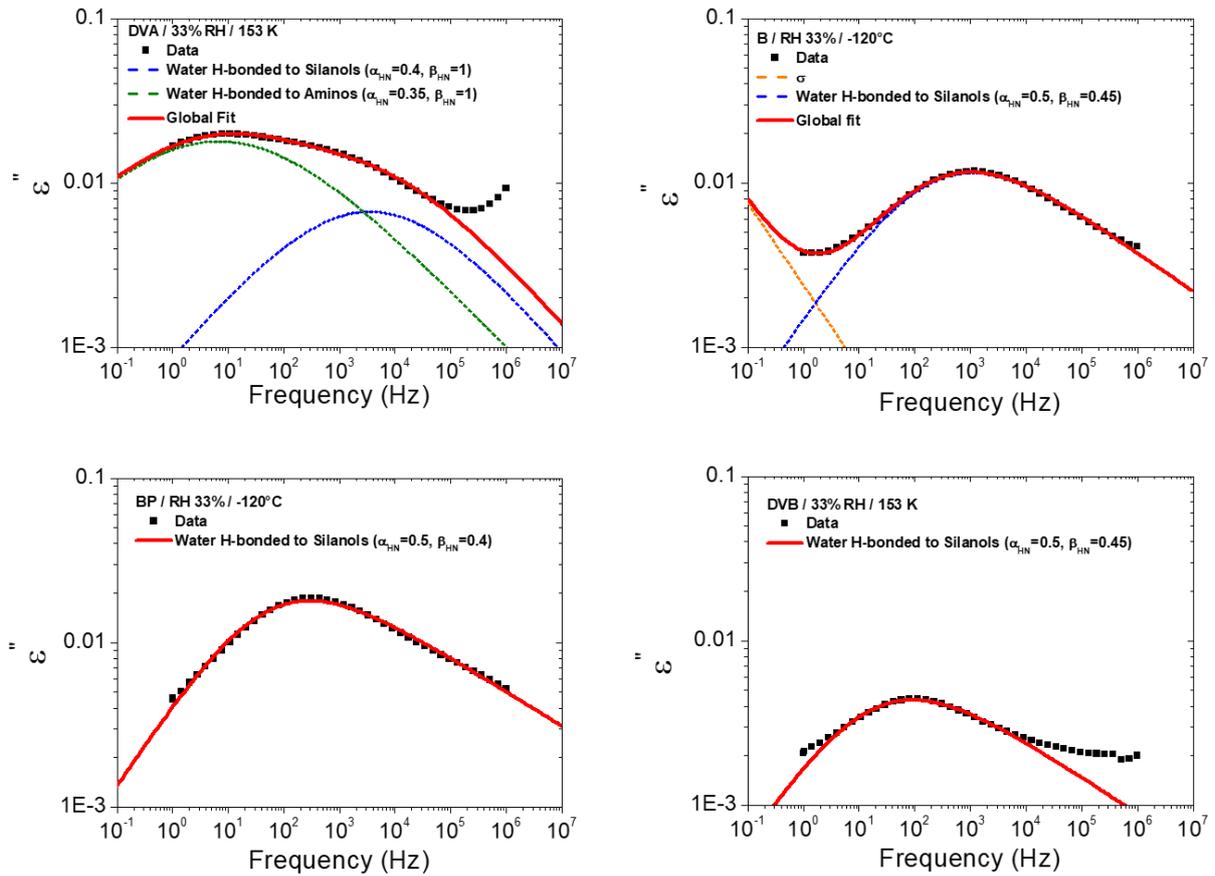

*Figure S9 Dielectric functions fitted to the experimental loss of water adsorbed in the four PMOs at 33% RH.*

## S11. Masterplot curves of the monolayer process in the five matrices

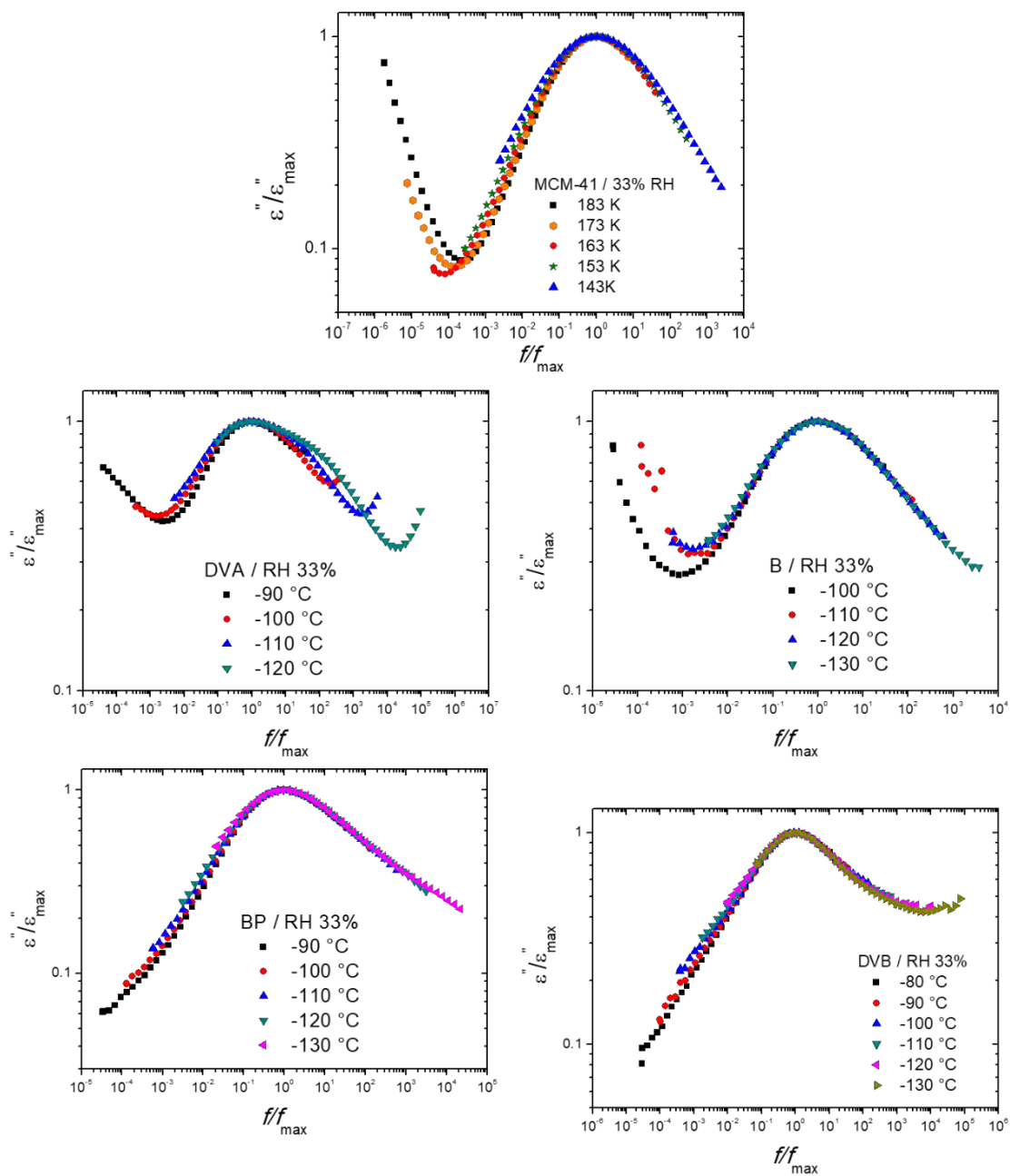

*Figure S10. Masterplot curves of the monolayer relaxation process in the five matrices.*

## S12. Temperature dependence of the different relaxation times for water adsorbed in mesoporous MCM-41 silicas at two RH.

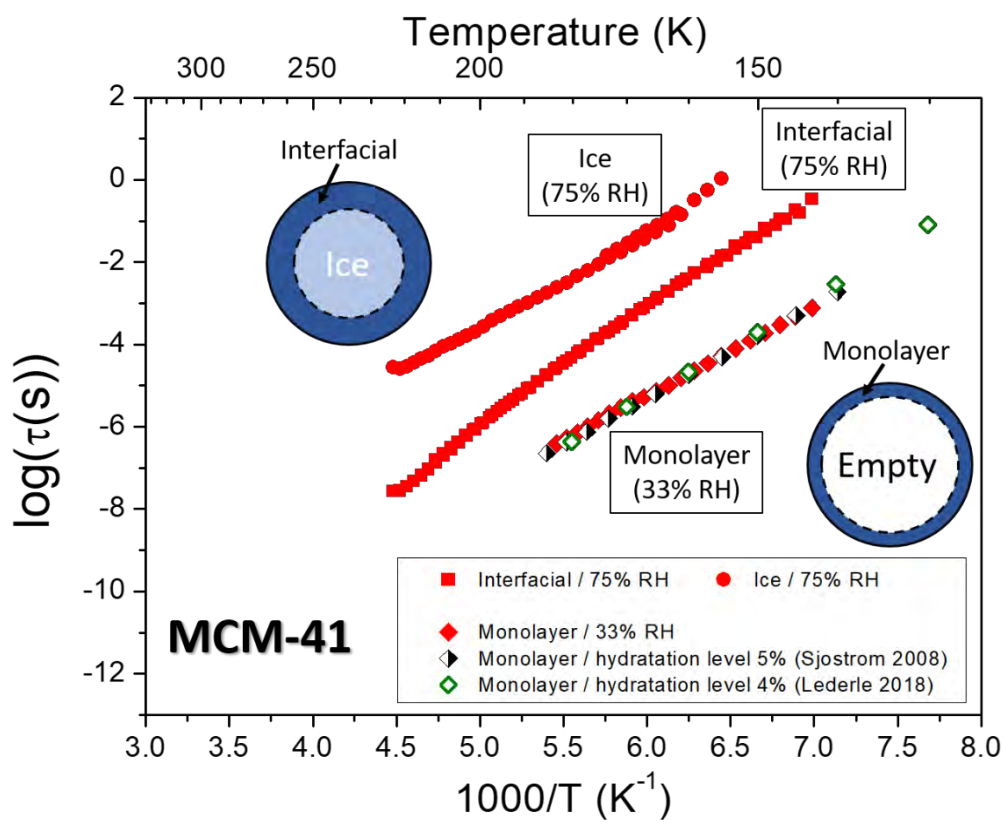

*Figure S11. Temperature dependence of relaxation times of the ice- and interfacial-processes of water confined into MCM-41 loaded at 75% RH; and monolayer-process of water confined into MCM-41 loaded at 33% RH. Loading at 75% RH is above capillary condensation and 33% RH is below. Additionally, data from literature[15,22] were plotted in order to strengthen the identification of processes.*

## S13. Highlighting crystallization and melting processes in SBA-15 loaded at 75% RH.

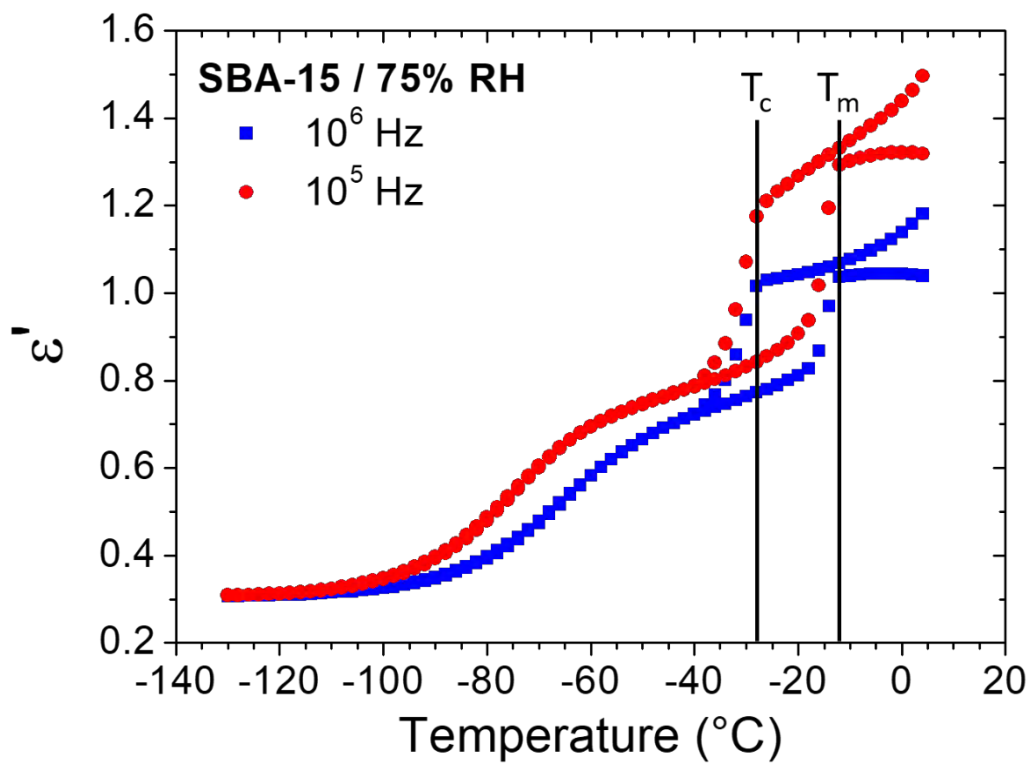

*Figure S12. Real part of the complex dielectric permittivity in isochronal representation, ε'(T), of confined water into SBA-15 loaded at 75% RH, i.e. above the capillary condensation.*

Figure S12 represents the real part of the complex dielectric permittivity in isochronal representation, ε'(T), for water confined in SBA-15 loaded at 75% RH, *i.e.* above the capillary-condensation, for two values of the frequency, $10^5$ and $10^6$ Hz. This representation has already demonstrated its ability to probe crystallization and melting processes[20,21]. Results clearly show a drop upon cooling at -28 °C and a jump upon heating at -12 °C, which are both frequency-independent. This is the signature of a crystallization process upon cooling at -28 °C and a melting process upon heating at -12 °C, confirming the capillary-filling of the pores. Contrariwise to crystallization/melting processes, a relaxation process is frequency-dependent as shown by the jump located at around -65 °C for the frequency $10^6$ Hz, whereas the jump is at around -75 °C for the frequency $10^5$ Hz.